\documentclass[twocolumn]{aastex631}

\usepackage[utf8]{inputenc} 
\usepackage{amsmath}
\usepackage{multirow}

\newcommand\teff{T$_{\rm{eff}}$}

\newcommand\msini{$M_p$ sin\textit{i}}

\newcommand\earthmass{$M_{\oplus}$}
\newcommand\earthradius{$R_{\oplus}$}
\newcommand\solmass{$M_{\odot}$}
\newcommand\solradius{$R_{\odot}$}

\hypersetup{linkcolor=violet,citecolor=purple}

\usepackage[inline]{enumitem}

\usepackage{listings}
\definecolor{lbcolor}{rgb}{0.9,0.9,0.9}
\DeclareTextSymbolDefault{\dh}{T1}

\received{October 1, 2023}
\revised{January 26, 2024}
\accepted{\today}

\shortauthors{Kanodia et al. 2024}
\submitjournal{AAS Journals}

\begin{document}

\title{Searching for Giant Exoplanets around M-dwarf Stars (GEMS) I: Survey Motivation}

\author[0000-0001-8401-4300]{Shubham Kanodia}
\affiliation{Earth and Planets Laboratory, Carnegie Institution for Science, 5241 Broad Branch Road, NW, Washington, DC 20015, USA}


\author[0000-0003-4835-0619]{Caleb I. Ca\~nas}
\altaffiliation{NASA Postdoctoral Fellow}
\affiliation{NASA Goddard Space Flight Center, 8800 Greenbelt Road, Greenbelt, MD 20771, USA }

\author[0000-0001-9596-7983]{Suvrath Mahadevan}
\affil{Department of Astronomy \& Astrophysics, 525 Davey Laboratory, The Pennsylvania State University, University Park, PA 16802, USA}
\affil{Center for Exoplanets and Habitable Worlds, 525 Davey Laboratory, The Pennsylvania State University, University Park, PA 16802, USA}

\author[0000-0001-6545-639X]{Eric B.\ Ford}
\affil{Department of Astronomy \& Astrophysics, 525 Davey Laboratory, The Pennsylvania State University, University Park, PA 16802, USA}
\affil{Center for Exoplanets and Habitable Worlds, 525 Davey Laboratory, The Pennsylvania State University, University Park, PA 16802, USA}
\affiliation{Institute for Computational and Data Sciences, The Pennsylvania State University, University Park, PA, 16802, USA}
\affiliation{Center for Astrostatistics, 525 Davey Laboratory, The Pennsylvania State University, University Park, PA, 16802, USA}

\author[0000-0001-5555-2652]{Ravit Helled}
\affil{Center for Theoretical Astrophysics \& Cosmology, University of Zurich, Winterthurerstr. 190, CH-8057 Zurich, Switzerland}

\author[0000-0002-8310-0554]{Dana E. Anderson}
\affil{Earth and Planets Laboratory, Carnegie Institution for Science, 5241 Broad Branch Road, NW, Washington, DC 20015, USA}

\author[0000-0001-7119-1105]{Alan Boss}
\affiliation{Earth and Planets Laboratory, Carnegie Institution for Science, 5241 Broad Branch Road, NW, Washington, DC 20015, USA}

\author[0000-0001-9662-3496]{William D. Cochran}
\affil{McDonald Observatory and Department of Astronomy, The University of Texas at Austin, USA}
\affil{Center for Planetary Systems Habitability, The University of Texas at Austin, USA}

\author[0000-0003-1439-2781]{Megan Delamer} 
\affil{Department of Astronomy \& Astrophysics, 525 Davey Laboratory, The Pennsylvania State University, University Park, PA 16802, USA}
\affil{Center for Exoplanets and Habitable Worlds, 525 Davey Laboratory, The Pennsylvania State University, University Park, PA 16802, USA}

\author[0000-0002-7127-7643]{Te Han} 
\affiliation{Department of Physics \& Astronomy, The University of California, Irvine, Irvine, CA 92697, USA}

\author[0000-0002-2990-7613]{Jessica E. Libby-Roberts}
\affil{Department of Astronomy \& Astrophysics, 525 Davey Laboratory, The Pennsylvania State University, University Park, PA 16802, USA}
\affil{Center for Exoplanets and Habitable Worlds, 525 Davey Laboratory, The Pennsylvania State University, University Park, PA 16802, USA}

\author[0000-0002-9082-6337]{Andrea S.J.\ Lin}
\affil{Department of Astronomy \& Astrophysics, 525 Davey Laboratory, The Pennsylvania State University, University Park, PA 16802, USA}
\affil{Center for Exoplanets and Habitable Worlds, 525 Davey Laboratory, The Pennsylvania State University, University Park, PA 16802, USA}

\author[0000-0002-8278-8377]{Simon M\"{u}ller}
\affil{Center for Theoretical Astrophysics \& Cosmology, University of Zurich, Winterthurerstr. 190, CH-8057 Zurich, Switzerland}

\author[0000-0003-0149-9678]{Paul Robertson}
\affiliation{Department of Physics \& Astronomy, The University of California, Irvine, Irvine, CA 92697, USA}

\author[0000-0001-7409-5688]{Gu\dh mundur Stef\'ansson}

\altaffiliation{NASA Sagan Fellow}
\affiliation{Department of Astrophysical Sciences, Princeton University, 4 Ivy Lane, Princeton, NJ 08540, USA}

\author[0009-0008-2801-5040]{Johanna Teske}
\affil{Earth and Planets Laboratory, Carnegie Institution for Science, 5241 Broad Branch Road, NW, Washington, DC 20015, USA}



\correspondingauthor{Shubham Kanodia}
\email{skanodia@carnegiescience.edu}

\begin{abstract}
Recent discoveries of transiting giant exoplanets around M-dwarf stars (GEMS), aided by the all-sky coverage of TESS, are starting to stretch theories of planet formation through the core-accretion scenario. Recent upper limits on their occurrence suggest that they decrease with lower stellar masses, with fewer GEMS around lower-mass stars compared to solar-type. In this paper, we discuss existing GEMS both through confirmed planets, as well as protoplanetary disk observations, and a combination of tests to reconcile these with theoretical predictions. We then introduce the \textit{Searching for GEMS} survey, where we utilize multi-dimensional nonparameteric statistics to simulate hypothetical survey scenarios to predict the required sample size of transiting GEMS with mass measurements to robustly compare their bulk-density with canonical hot-Jupiters orbiting FGK stars. Our Monte-Carlo simulations predict that a robust comparison requires about 40 transiting GEMS (compared to the existing sample of $\sim$ 15) with 5-$\sigma$ mass measurements. Furthermore, we discuss the limitations of existing occurrence estimates for GEMS, and provide a brief description of our planned systematic search to improve the occurrence rate estimates for GEMS.
\end{abstract}


\section{Introduction} \label{sec:intro}

M-dwarfs are the most common type of star in the Galaxy \citep{reid_low-mass_1997, henry_solar_2006, reyle_10_2021}. The M-dwarf spectral type spans almost an order of magnitude in mass ranging from $\sim$ 0.08~\solmass~to $\sim$ 0.6 \solmass, and about 2600 K to 4000 K in effective temperature \citep{pecaut_intrinsic_2013}. Compared to solar-type stars, these low mass stars are expected to have correspondingly lower mass protoplanetary disks \citep{andrews_mass_2013, pascucci_steeper_2016}, and longer Keplerian orbital timescales (at a fixed distance). The combination of these factors is theorized to make it difficult to form giant planets around these stars in protoplanetary disks (Class II) under the core-accretion formation paradigm. Under this paradigm, traditionally it has been thought that a rocky heavy-element core of roughly $\sim 10$~\earthmass~ must first form, which is then followed by runaway gaseous accretion to rapidly accumulate a massive gaseous envelope\footnote{See \cite{helled_giant_2014, dawson_origins_2018, fortney_hot_2021, helled_planet_2021} for comprehensive reviews on giant planet formation.} \citep{mizuno_formation_1980, pollack_formation_1996}. Early studies showed that due to the lower disk masses and longer orbital timescales, the formation of a protoplanet massive enough to initiate runaway gaseous accretion would take too long with respect to the lifetime of the gas (primarily H/He) in protoplanetary disks \citep{laughlin_core_2004, ida_toward_2005}. An alternative rapid formation mechanism has been proposed in the form of gravitational instability \citep[GI;][]{boss_giant_1997, boss_rapid_2006}, which takes place during the proto-stellar phase (Class 0 or I disk) of massive disks when the star is still embedded in a molecular cloud \citep{lada_star_1987, dauphas_perspective_2011}.

While it is estimated that M-dwarfs host multiple small terrestrial planets on average \citep{dressing_occurrence_2015, hardegree-ullman_kepler_2019, hsu_occurrence_2020}, the occurrence of giant planets around these types of stars is more uncertain due to their rarity \citep{endl_exploring_2006, johnson_giant_2010, maldonado_connecting_2019, schlecker_rv-detected_2022, gan_occurrence_2023, bryant_occurrence_2023}.  Attempts to understand the occurrence of giant exoplanets around M-dwarf stars (GEMS) have traditionally been limited to radial velocity (RV) surveys \citep{endl_exploring_2006, johnson_giant_2010, maldonado_connecting_2019, sabotta_carmenes_2021, schlecker_rv-detected_2022, pinamonti_hades_2022}, since M-dwarfs accounted for only a minor fraction of the target stars observed by the Kepler mission \citep{borucki_kepler_2010}. This has recently started to change with NASA's Transiting Exoplanet Survey Satellite (TESS), and its all-sky coverage that includes millions of bright M-dwarfs amenable to RV follow-up \citep{ricker_transiting_2014, muirhead_catalog_2018}. Despite the low predicted occurrence of GEMS, enough M-dwarf host stars have been observed within the first few TESS cycles that attempts have been made to characterize the occurrence rate of transiting GEMS \citep{gan_occurrence_2023, bryant_occurrence_2023}.  However in subsequent sections, we discuss how these recent investigations are just a first step and motivate a more detailed analysis and characterization of the TESS detection sensitivity for GEMS, and subsequent estimation of their occurrence.

In this manuscript, we present the motivation for our \textit{Searching for GEMS} survey, where in Section \ref{sec:evidence} we present the existing GEMS in planet samples, and protoplanetary disks. In Section \ref{sec:formation}, we discuss predictions from different formation and population synthesis models, and observational tests to distinguish between formation mechanisms. Next, in Section \ref{sec:outline} we introduce our \textit{Searching for GEMS} survey, its requirements, motivations and provide a brief outline, before summarizing our work in Section \ref{sec:summary}. Preliminary results and trends seen in these existing samples will be discussed and evaluated in part II of this work.

\section{GEMS}\label{sec:evidence}

\begin{figure}
\centering
\includegraphics[width=\columnwidth]{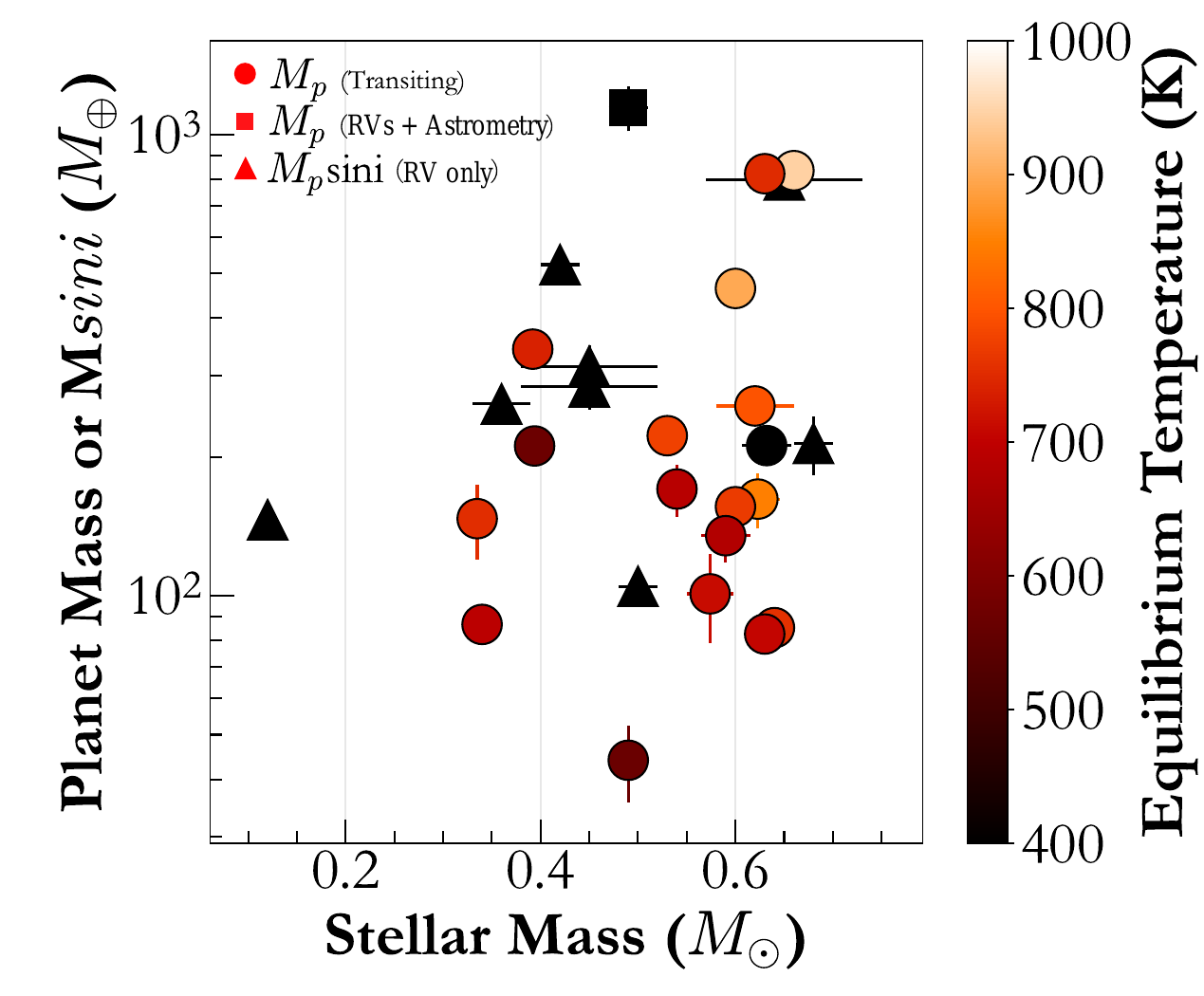}
\caption{Planetary mass plotted as a function of stellar mass for GEMS. The transiting planets have true mass measurements, whereas \msini{} is plotted for the non-transiting ones, with the exception of GJ 463~b which has a true mass measurement of $\sim 1140$ \earthmass{} (or $\sim$ 3.6 $M_J$) from astrometry \citep{sozzetti_dynamical_2023}. }
\label{fig:gems_masses}
\end{figure} 

\subsection{As Exoplanets}

In \autoref{fig:gems_masses} we show the transiting and non-transiting GEMS as queried from the NASA Exoplanet Archive on 2023 June 7 \citep{akeson_nasa_2013, PSCompPars} resulting in 10 non-transiting (with \msini{} $>$ 100 \earthmass) and 18 transiting GEMS (with planetary radius $\gtrsim$ 8 \earthradius) orbiting stars with \teff{} $\lesssim$ 4000 K, with precise masses $>$ 3-$\sigma$ (\autoref{tab:InputSample}). In addition, there have been some detections of GEMS with microlensing \citep{suzuki_exoplanet_2016}, and direct imaging \citep{lannier_massive_2016, nielsen_gemini_2019} which contribute to statistical analysis; this will especially be true with hundreds of expected detections from Roman Space Telescope \citep{penny_predictions_2019}. While the transiting GEMS have largely been clustered around early M-dwarfs (M0-M2), there have been recent discoveries of four mid M-dwarf GEMS --- TOI-5205~b \citep{kanodia_toi-5205b_2023}, TOI-3235~b \citep{hobson_toi-3235_2023}, TOI-519~b \citep{kagetani_mass_2023}, and TOI-4860 b \citep{almenara_toi-4860_2023, triaud_m_2023}. The mid M-dwarf GEMS, in addition to recent RV detections around mid-to-late M-dwarfs \citep[e.g.][]{morales_giant_2019, quirrenbach_carmenes_2022}, defy expectations from population synthesis models in protoplanetary disks and break the inferred mass-budget\footnote{An extreme example of this was the discovery of a 5 $M_J$ planet around a 25 $M_J$ brown dwarf, which likely formed through gravitational instability or fragmentation \citep{lodato_constraints_2005}.}. 

We also note that while most of these transiting GEMS fall in the canonical hot-Jupiter parameter space based on their typical orbital periods of $<$ 10 days, it is erroneous to classify them as hot. Because of their cooler and lower-mass M-dwarf hosts, the transiting GEMS are not `hot', with equilibrium temperatures $<$ 1000 K, and hence not expected to be inflated like hot-Jupiters \citep{weiss_mass_2013, dawson_origins_2018, thorngren_bayesian_2018}. Furthermore, their $a/R_*$ are $\gtrsim 10$, and hence these objects may not be tidally locked or tidally circularized like hot-Jupiters. Thus, transiting GEMS likely have different planetary properties (both bulk and atmospheric) than hot-Jupiters, and should not be classified as such.

\begin{deluxetable*}{cccccccccccc}
    \tabletypesize{\footnotesize}
    \tablecaption{Transiting Giant Exoplanets around M-dwarf Stars (GEMS) with $R_p$ $\gtrsim$ 8 \earthradius, and \teff $<$ 4000 K to within 1-$\sigma$.  \explain{Add uncertainties for Teff} \label{tab:InputSample}}
    \tablehead{
    \colhead{Planet Name} & \colhead{Pl. Mass} & \colhead{Pl. Radius} & \colhead{Orbital Period} & \colhead{Insolation} & \colhead{\teff} & \colhead{St. Mass} & \colhead{Distance} & \colhead{This} &  \colhead{References} \\
    \colhead{} & \colhead{\earthmass{}} & \colhead{\earthradius{}} & \colhead{d} & \colhead{$S_{\oplus}$}  & \colhead{K} & \colhead{\solmass} & \colhead{pc} & \colhead{Survey}}
    \startdata
TOI-3984 A b    &  44.00$^{+8.73}_{-7.99}$      &  7.90$^{+0.24}_{-0.24}$       &  4.3533       &  17.19 $\pm$ 1.98     &  3480 $\pm$ 90 &  0.49 $\pm$ 0.02      &  108 & \textsuperscript{$\bigstar$}& \textit{a} \\
TOI-3629 b      &  82.64 $\pm$ 6.36     &  8.29 $\pm$ 0.22      &  3.9366       &  41.44 $\pm$ 4.82     &  3870 $\pm$ 90 &  0.63 $\pm$ 0.02      &  130 & \textsuperscript{$\bigstar$} & \textit{b, i} \\
TOI-4860 b\textsuperscript{\textdaggerdbl}      &  86.70 $\pm$ 1.90     &  8.58 $\pm$ 0.30      &  1.5228       &  38.64 $\pm$ 3.41     &  3255 $\pm $49 &  0.34 $\pm$ 0.01      &  80 & - & \textit{c}\\
TOI-5344 b      &  135.00 $\pm$ 17.00   &  9.80 $\pm$ 0.50      &  3.7926       &  35.58 $\pm$ 3.05     &  3757 $\pm$ 51 &  0.59$^{+0.03}_{-0.02}$       &  136 & \textsuperscript{$\bigstar$}& \textit{d, i} \\
HATS-75 b       &  156.05 $\pm$ 12.40   &  9.91 $\pm$ 0.15      &  2.7887       &  58.29 $\pm$ 0.74     &  3790 $\pm$ 6 &  0.60 $\pm$ 0.01      &  195 & - & \textit{e}\\
TOI-1899 b      &  211.71$^{+13.76}_{-13.54}$   &  11.11$^{+0.36}_{-0.34}$      &  29.0903      &  3.06 $\pm$ 0.33      &  3909 $\pm$ 90 &  0.63$^{+0.03}_{-0.03}$       &  128 & \textsuperscript{$\bigstar$}&  \textit{f} \\
HATS-6 b        &  101.00 $\pm$ 22.00   &  11.19 $\pm$ 0.21     &  3.3253       &  42.73 $\pm$ 2.09     &  3724 $\pm$ 18 &  0.57$^{+0.02}_{-0.03}$       &  169 & - &  \textit{g} \\
TOI-3714 b      &  222.48 $\pm$ 9.53    &  11.32 $\pm$ 0.34     &  2.1548       &  60.02 $\pm$ 6.53     &  3660 $\pm$ 90 &  0.53 $\pm$ 0.02      &  113 & \textsuperscript{$\bigstar$} &  \textit{b, i} \\
TOI-3235 b      &  211.35 $\pm$ 7.95    &  11.40 $\pm$ 0.49     &  2.5926       &  17.47 $\pm$ 2.19     &  3196 $\pm$ 67 &  0.39 $\pm$ 0.02      &  73 & - & \textit{h} \\
TOI-519 b       &  147.00 $\pm$ 27.00   &  11.55 $\pm$ 0.33     &  1.2652       &  52.98 $\pm$ 4.43     &  3322 $\pm$ 49 &  0.34 $\pm$ 0.01      &  115 & - & \textit{j, i} \\
HATS-74 A b*     &  464.03 $\pm$ 44.50   &  11.57 $\pm$ 0.23     &  1.7319       &  108.46 $\pm$ 4.08    &  3777 $\pm$ 10 &  0.60 $\pm$ 0.01      &  299 & - & \textit{e} \\
TOI-5205 b      &  342.70$^{+17.70}_{-16.80}$   &  11.60 $\pm$ 0.30     &  1.6308       &  49.24 $\pm$ 4.24     &  3430 $\pm$ 54 &  0.39 $\pm$ 0.01      &  87 & \textsuperscript{$\bigstar$}&  \textit{k} \\
TOI-5293 A b    &  170.45$^{+21.84}_{-21.86}$   &  11.89$^{+0.44}_{-0.41}$      &  2.9303       &  37.70 $\pm$ 4.16     &  3586 $\pm$ 90 &  0.54 $\pm$ 0.02      &  161 & \textsuperscript{$\bigstar$}& \textit{a} \\
TOI-3757 b      &  85.30$^{+8.80}_{-8.70}$      &  12.00$^{+0.40}_{-0.50}$      &  3.4388       &  54.81 $\pm$ 3.78     &  3913 $\pm$ 56 &  0.64 $\pm$ 0.02      &  181 & \textsuperscript{$\bigstar$}&  \textit{l} \\
Kepler-45 b*     &  162.09 $\pm$ 22.25   &  12.11$^{+0.34}_{-0.45}$      &  2.4552       &  85.88 $\pm$ 6.11     &  3950 $\pm$ 40 &  0.62$^{+0.02}_{-0.02}$       &  382 & - &\textit{m} \\
HATS-76 b*       &  835.57 $\pm$ 28.29   &  12.09 $\pm$ 0.35     &  1.9416       &  131.80 $\pm$ 5.44    &  4016 $\pm$ 17 &  0.66 $\pm$ 0.02      &  386 & - & \textit{e} \\
TOI-4201 b      &  823.00$^{+21.00}_{-20.00}$   &  13.11 $\pm$ 0.45     &  3.5819       &  52.83 $\pm$ 4.49     &  3920 $\pm$ 50 &  0.63 $\pm$ 0.02      &  188 & \textsuperscript{$\bigstar$}& \textit{n, o, i} \\
NGTS-1 b*\textsuperscript{\textdagger}        &  258.08$^{+20.98}_{-23.84}$   &  14.91$^{+6.84}_{-3.70}$      &  2.6473       &  67.26 $\pm$ 19.55    &  3916 $\pm$ 67 &  0.62$^{+0.02}_{-0.06}$       &  218 & - &\textit{p} \\
    \enddata
    \tablenotetext{\textsuperscript{$\bigstar$}}{ GEMS discovered through this survey}
    \tablenotetext{*}{GEMS with distance $>$ 200 pc, and hence will not be included in the occurrence rate sample. We note that there exists HATS-77~b around a dwarf star with \teff{} of 4071 K, which is ostensibly an M-dwarf host \citep{jordan_hats-74ab_2022}. However, since it the system is $>$ 200 pc, it is not included in the statistical sample either}
    \tablenotetext{$\textdaggerdbl$}{TOI-4860~b was also confirmed in a separate publication by \cite{triaud_m_2023}.}
    \tablenotetext{$\textdagger$}{NGTS-1 b is not shown in the plots in this manuscript since it has an imprecise radius to due its grazing transit.}
    \tablenotetext{}{\textit{a)} \cite{canas_toi-3984_2023}, \textit{b)} \cite{canas_toi-3714_2022}, \textit{c)} \cite{almenara_toi-4860_2023}, \textit{d)} Han et al. (in prep.), \textit{e)} \cite{jordan_hats-74ab_2022}, \textit{f)} \cite{canas_warm_2020, lin_unusual_2023}, \textit{g)} \cite{hartman_hats-6b_2015}, \textit{h)} \cite{hobson_toi-3235_2023}, \textit{i)} \cite{hartman_toi_2023}, \textit{j)} \cite{kagetani_mass_2023},  \textit{k)} \cite{kanodia_toi-5205b_2023}, \textit{l)} \cite{kanodia_toi-3757_2022}, \textit{m)} \cite{johnson_characterizing_2012}, \textit{n)} \cite{delamer_toi-4201_2023}, \textit{o)} \cite{gan_massive_2023}, \textit{p)} \cite{bayliss_ngts-1b_2018}}
\end{deluxetable*}

\subsection{As structure in Protoplanetary Disks}

In addition to the direct detection of exoplanets, the high-resolution and high-contrast imaging of disks has enabled the detection of structures and gaps in Class II protoplanetary disks. While there are a number of physical mechanisms that have been proposed to explain these structures, the presence of young proto-planets has been the topic of recent investigations \citep{dong_observational_2015}.  For example, \cite{van_der_marel_stellar_2021} identify potential trends between the incidence of structured disks and stellar mass that can be accounted for by massive Jovian-sized exoplanets.  While they note a reduction in the prevalence of structured (rings or transition) with lower stellar masses, roughly 1 -- 10 $\%$ of their $>$ 100 disks around such stars require the presence of a $>$ 1 $M_J$ planet to explain the features present. These trends were extended to very low mass stars (VLMS) and brown dwarfs using high-angular resolution observations from ALMA, which suggested the presence of substructures in a fraction of these disks \citep{pinilla_first_2022}. \cite{zhang_substructures_2023} perform a similar analysis and compare the incidence of structures in disks in the Taurus star-forming region to infer the occurrence of protoplanets as a function of stellar mass and orbital-separation.

Additionally, \cite{curone_giant_2022} performed a detailed analysis of the disk CIDA 1 around a $\sim$ 0.2 \solmass{} star and the substructures present therein, and indicate a planet $>$ 1.4 $M_J$ at $\sim$ 10 AU being responsible for the observed morphology. Similarly \cite{long_large_2023} note the presence of two gaps in the disk around an M3.5 star J04124068+2438157, which they attribute to a Saturn-mass planet at $\sim$ 90 AU.

To summarize, there is observational evidence regarding the existence of GEMS at large separations in young (Class II, i.e., $<$ 10 Myr) systems, and at closer separations in mature stellar systems. This will be further bolstered by the addition of hundreds of GEMS projected to be astrometrically detected through Gaia DR4 \citep{sozzetti_astrometric_2014, perryman_astrometric_2014}.

\section{\textit{Searching for GEMS} Survey Motivation}\label{sec:formation}

\subsection{Forming GEMS}
\subsubsection{Formation during protoplanetary phase}

\begin{figure}
\centering
\includegraphics[width=\columnwidth]{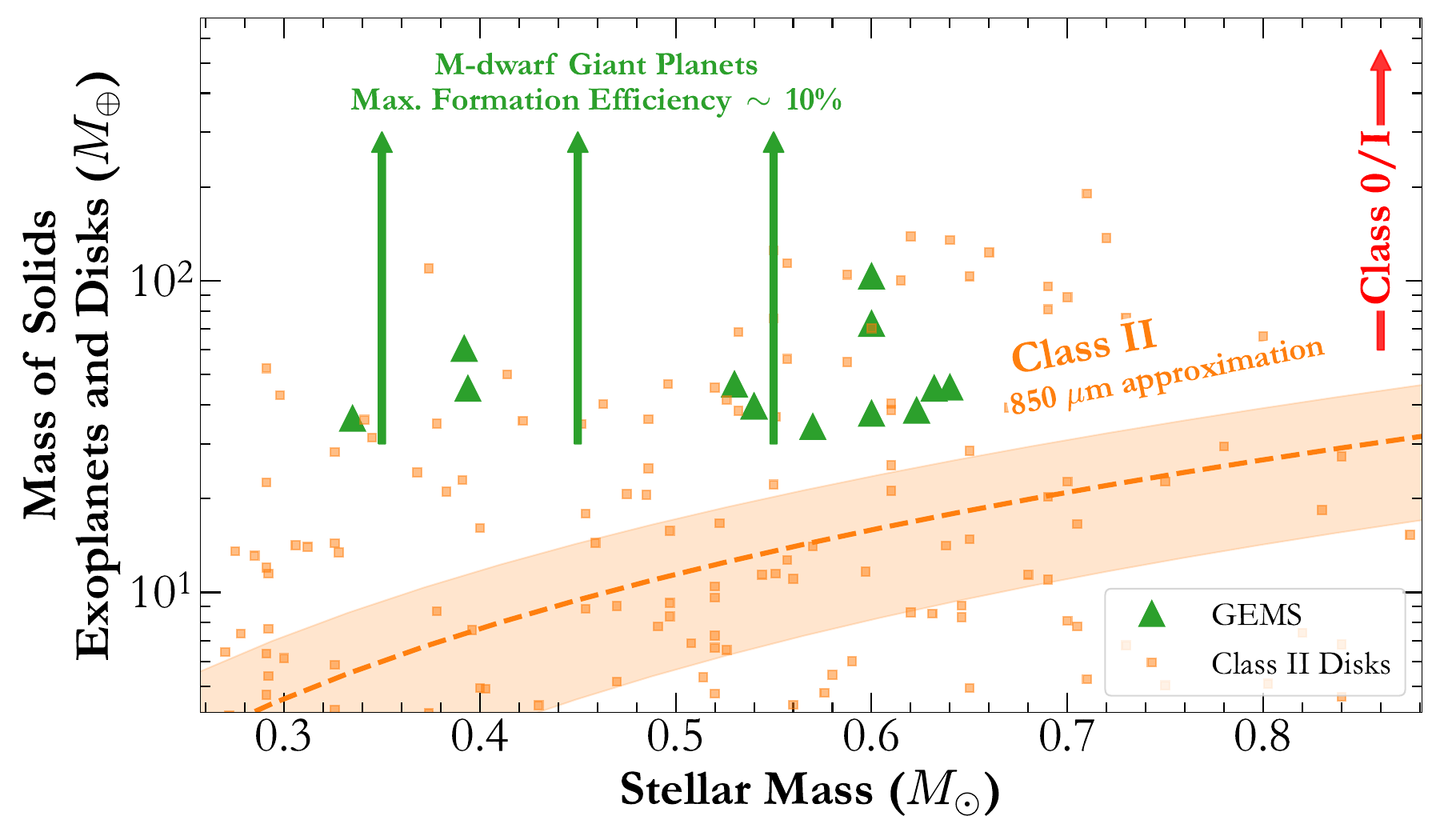}
\caption{Green triangles show an estimate of the heavy-element content ($M_Z$) of GEMS from planetary interior models  \citep{thorngren_mass-metallicity_2016} compared to the Class II disk dust mass ($M_d$) estimates as orange squares \citep{manara_demographics_2022} and the median (and 1-$\sigma$) trend seen in the Lupus sample \citep{ansdell_alma_2016}. The green arrows show the approximate expected disk dust masses required to form the green triangles assuming 10\% formation efficiency \citep{liu_growth_2019}. The red line shows the range of dust masses for Class 0 and I disks from \cite{tychoniec_dust_2020}. \textbf{Takeaway:} The formation of GEMS necessitates disks with many 100s of \earthmass{} of heavy-elements. This could be achieved by anomalously massive Class II disks with underestimated dust masses, or Class 0/I disks.}
\label{fig:dustmasses}
\end{figure}

Under the core-accretion paradigm \citep[][]{mizuno_formation_1980} --- ignoring migration and transport of material across the disk --- a massive solid core is formed by accretion inside the protoplanet's feeding zone \citep[roughly a few Hill radii;][]{ida_toward_2004_1, alibert_models_2005}. Once this zone is depleted, the planet reaches isolation mass where its accretion slows down and takes place primarily through the accretion of gas on to its envelope up until it reaches crossover mass where the mass of solids is roughly equal to the mass of gas (planetary mass of $\sim$ 30 \earthmass; heavy-element mass of $\sim 10-20$ \earthmass). If this threshold is reached when the disk still contains its gas, it initiates exponential runaway gaseous accretion where a Neptune sized planet accretes enough gas to become a gas giant.  We note the caveats that recent studies have suggested that the energy released during accretion of the solid core might delay runaway gaseous accretion \citep{venturini_jupiters_2020, kessler_interplay_2023}, such that the runaway phase only occurs much later at a planet mass of $\sim 100$ \earthmass~ with a heavy-element mass of 20--30 \earthmass. On the other hand, simulations have also shown the feasibility of forming giant planets with lower mass cores of $\sim$ 4 \earthmass{} in lower surface density disks, albeit over much longer times-scales \citep{movshovitz_formation_2010}. Therefore predictions for solid core mass are dependent on models and initial conditions assumed. Yet GEMS, especially those around mid-to-late M-dwarfs, continue to present a challenge for theories of planet formation during the protoplanetary phase, i.e., core-accretion \citep{ida_toward_2005, liu_super-earth_2019,miguel_diverse_2020, burn_new_2021}.

Summarizing briefly, there are two main reasons for this difficulty in forming GEMS in the protoplanetary phase with core accretion. The first is the  the dust mass\footnote{Given the greater than order of magnitude uncertainties associated with the estimates for these quantities, in this manuscript we use the terms dust, heavy-elements, solids and metals interchangeably.} budget of the disk available to form giant planets. Based on estimates of exoplanet heavy-element content ($M_Z$) and efficiency of planet formation, \cite{kanodia_toi-5205b_2023} show how median Class II disks might not have enough dust mass to form GEMS. In \autoref{fig:dustmasses} we compare the estimated heavy-element content ($M_Z$) for GEMS from planetary interior models \citep{thorngren_mass-metallicity_2016}, with disk dust masses ($M_d$) and show that assuming an optimistic planet formation efficiency of 10\% through pebble accretion \citep{liu_growth_2019} necessitates disks with many 100s of \earthmass{} of heavy-elements. We note the caveat here that recent giant planet interior models tend to predict a lower heavy-element content when incorporating results from newer equations of state, results from the Solar-system gas giants, and atmospheric metallicity estimates \citep{miguel_jupiters_2022, miguel_interior_2023, muller_synthetic_2021, muller_towards_2022}, which might alleviate some of this mass deficit. The second is the timescale of formation of a solid massive core to initiate runaway gaseous accretion: the crossover mass must be attained before the gas in the disk is accreted on to the host star. \cite{laughlin_core_2004} show that the lower disk masses (and subsequently surface density) coupled with longer Keplerian timescales (also due to lower host stellar masses) mean that planetary cores around M-dwarfs at 5 AU might take too long to form relative to the lifetime of the disks. 


\textit{Underestimated disk dust masses}: During the protoplanetary phase, i.e. Class II disks, the dust mass\footnote{See \cite{miotello_setting_2022} for a review of disk dust mass estimation.} is typically estimated based on continuum flux measurements of mm sized dust particles, either extrapolating from measurements at 850 $\mu$m \citep{hildebrand_determination_1983}, or fitting SEDs to multi-wavelength data \citep{pinte_probing_2008}. These flux-to-mass estimates break down if the continuum emission is optically thick \citep{eisner_protoplanetary_2018, rilinger_determining_2023, xin_measuring_2023}, or in the presence of gaps or rings in the disk \citep{liu_underestimation_2022}, leading to underestimation of the dust masses by 3 -- 10x. 

More fundamentally, the dust mass in mm sized particles in ALMA measurements does not represent the true primordial mass budget, with formation already underway locking up the dust in $>$ mm sized particles \citep{greaves_have_2010, najita_mass_2014}, as well as depletion due to radial drift \citep{appelgren_disc_2023}. If the accretion processes required to form a massive core (to initiate runaway gaseous accretion) started earlier in the protoplanetary stage, the true mass budget available for planet formation would be larger than the 850 $\mu$m mass estimates.


\textit{Alleviating the timescale problem}: The timescale problem could potentially be circumvented by pebble accretion, which is a much faster process than planetesimal accretion\footnote{Though population synthesis by \cite{liu_super-earth_2019} still struggled to form giant planets around low-mass stars through pebble-accretion.} \citep{lambrechts_rapid_2012, savvidou_how_2023}. Furthermore, new theories suggest that some young disks could concentrate solids in their mid-plane through self-gravitating spiral waves, thereby hastening the formation of a core massive enough to reach the runaway stage \citep{haghighipour_gas_2003, baehr_filling_2023}. Lastly, studies of large disk clusters have shown that the low mass M-dwarf disks tend to have longer lifetimes than those around more massive stars \citep{pfalzner_most_2022}, with some extreme examples existing in the form of so-called `Peter-Pan' disks lasting many 10s of Myr \citep{flaherty_planet_2019, silverberg_peter_2020, coleman_peter_2020, wilhelm_exploring_2022}. Together these factors might ease the timescale problem against the formation of GEMS.

Thus, to summarize, while it has traditionally been thought to have been difficult to form GEMS through core-accretion during the protoplanetary disk phase, this formation pathway cannot be ruled out given the dependence on numerous poorly understood disk parameters.

\subsubsection{Formation during protostellar phase}

An alternative to core accretion is the formation of GEMS in the protostellar phase, i.e., in Class 0, I disks (or protostars) through gravitational instability \citep[GI;][]{kuiper_origin_1951, cameron_physics_1978, boss_giant_1997}. Under this mechanism, massive disks can start fragmenting and forming dense self-gravitating clumps as precursors to giant planets\footnote{See \cite{durisen_gravitational_2007} for a review of GI.}. Gravitational collapse is prevented close to the star due to the faster rotation period and thermal pressure (higher temperatures). \cite{boss_rapid_2006, boss_formation_2011} show how GI can form giant planets around M-dwarfs with mass 0.1 and 0.5 \solmass{} at large separations, which is then followed by migration induced by the spiral arms \citep{boss_orbital_2023}. \cite{mercer_planet_2020} run smoothed particle hydrodynamics (SPH) simulations to show that the GI formation of GEMS necessitates disk-to-star mass ratios between $\sim 0.3$ to 0.6. \cite{haworth_massive_2020} presented a range of SPH models for massive disks around M-dwarf stars to show the difference between disk-to-star mass ratios for disks that remain axisymmetric, produce spiral arms or fragment into clumps. \cite{boss_forming_2023} perform GI-based population synthesis to quantify the frequency with which GEMS can be formed around a range of stellar masses ranging from 0.1 to 0.5 \solmass{} and disk-to-star mass ratios of 0.05 to 0.3. These mass ratios are consistent with those seen in Class 0/I protostellar samples from VLA \citep{tychoniec_vla_2018, tychoniec_dust_2020, xu_testing_2022, fiorellino_mass_2023}.

\subsection{Observationally distinguishing between formation mechanisms}
In this subsection, we consider and establish a few potential mechanisms to observationally distinguish between forming GEMS via core-accretion vs. gravitational instability.

\subsubsection{Stellar metallicity trends}\label{sec:stellarmetallicity}
Since the discovery of the first few gas giant exoplanets studies have shown that the occurrence of these planets around FGK stars is positively correlated with stellar metallicity \citep{gonzalez_stellar_1997, santos_metal-rich_2001, fischer_planet-metallicity_2005, wang_revealing_2015, petigura_california-keplersurvey_2018, narang_properties_2018, osborn_investigating_2020}. This positive metallicity trend agrees well with predictions from the core-accretion model of formation \citep{ida_toward_2004_2, matsumura_n-body_2021}, where higher metallicity stars correspond to higher metallicity disks (higher solid surface density) and translates into faster solid core formation\footnote{Here we simplistically assume uniform disk metallicity, which is likely inaccurate given the possibility of dust settling, local metallicity gradients and bumps.}. This explanation of the metallicity trend is further aided by observational evidence that higher metallicity disks tend to live longer \citep{yasui_lifetime_2009, yasui_short_2010}, with the higher opacity potentially shielding the disk against photoevaporation and reducing disk accretion \citep{yasui_low-metallicity_2021}. Overall, numerous simulations and observational studies have shown that the core accretion paradigm of planet formation favours higher metallicity host stars.

On the contrary, simulations suggest that lower metallicity molecular clouds should favour gravitational fragmentation \citep{matsukoba_protostellar-disc_2022, elsender_statistical_2021}, while GI is largely agnostic of the disk metallicity \citep{boss_stellar_2002}. Put simply, in the optically thick environment of the disk, higher stellar metallicity leads to higher opacities (due to more free electrons in the disk) that can slow down the cooling timescales. However, \cite{boss_stellar_2002} showed that at the typical separations at which GI occurs, the temperature equilibrates much faster than the dynamical timescales and a factor of few ($\pm$ 0.5 dex) change in metallicity does not make an appreciable difference in the efficiency of GI. 

This difference in stellar metallicity dependence is indeed seen as one of the cleaner ways to distinguish between the two mechanisms of giant planet formation given a large enough sample, which also sidesteps the numerous complications associated with other methods such as atmospheric chemistry \citep{venturini_jupiters_2020, molliere_interpreting_2022}. This difference has been noticed in the observed sample of giant planets and brown dwarfs orbiting FGK stars around 4 -- 10 $M_J$ \citep{santos_observational_2017, schlaufman_evidence_2018, narang_properties_2018}, with lower mass giant planets ($\sim$ 1--4 $M_J$) preferring more metal rich stars than the more massive giant planets ($>$ 10 $M_J$). This feature is hypothesized to be explained by prevalence of core-accretion formed planets below this transition, and GI formed objects (planets or brown dwarfs) above it.

\paragraph{Complexities in M dwarf metallicity determination}\label{sec:complexities}

While studies have attempted to extend the positive metallicity dependence of giant planets to the M-dwarf spectral type \citep{johnson_california_2010, maldonado_connecting_2019, gan_toi-530b_2022}, with their cooler stellar atmospheres, M-dwarfs are blanketed with molecular features that cause complexities in metallicity determination. The molecular features complicate continuum estimation \citep{pineda_using_2013} and cause inaccuracies in line profile measurement due to line blending, especially in the optical, where the spectra is dominated by TiO as the dominant source of opacity \citep{kirkpatrick_standard_1991, jorgensen_effects_1994}. \cite{rains_characterization_2021} discuss how the errors in the TiO line lists \citep{mckemmish_exomol_2019} manifest in discrepancies in model atmospheres. These issues are further exacerbated by low S/N observations in the optical due to the low effective temperatures and luminosities of these stars. 

Recently, the introduction of high-resolution near-infrared spectrographs has helped ameliorate this situation as metallicity determination has moved from photometric relations \citep{bonfils_metallicity_2005, schlaufman_physically-motivated_2010, neves_metallicity_2013} and spectral indices \citep{terrien_h-band_2012, mann_prospecting_2013, kuznetsov_characterization_2019} to empirical spectral comparison methods \citep{yee_precision_2017, stefansson_sub-neptune-sized_2020}. With increased sophistication and accuracy for stellar atmospheric models, spectral synthesis techniques \citep{bean_metallicities_2006, woolf_metallicity_2005, lindgren_metallicity_2016, souto_stellar_2020, souto_metallicity_2021, souto_detailed_2022, tabernero_steparsyn_2022,  iyer_sphinx_2022} and data driven approaches \citep{antoniadis-karnavas_odusseas_2020, birky_temperatures_2020, passegger_carmenes_2020} have demonstrated a further improvement. Additionally, studies are now moving towards elemental abundance measurements for M dwarfs \citep{veyette_physically_2017, souto_chemical_2017, maldonado_hades_2020, ishikawa_elemental_2021, shan_carmenes_2021}, similar to those for FGK stars \citep{hinkel_comparison_2016, teske_metal-rich_2019, wilson_influence_2021}. Uniform high-resolution H-band observations with APOGEE of all confirmed exoplanets and TOIs through SDSS V \citep{kollmeier_sdss-v_2017} may offer the opportunity to utilize existing techniques that have been demonstrated to work well for M-dwarf metallicities.

\subsubsection{Planet atmospheric composition}
Studies have suggested that the metallicity and composition of the atmospheres of giant planets could hold clues to their formation and evolution history, ranging from the mechanisms and location of formation, and the subsequent migration if applicable \citep[][ and numerous others]{helled_metallicity_2010, oberg_effects_2011, madhusudhan_toward_2014, knierim_constraining_2022, dash_linking_2022}. Further to the above discussions, studies have suggested potential differences in planetary atmospheres formed through core-accretion vs GI \citep{hobbs_molecular_2022}. However, this initial picture is likely to be complicated by location of formation and subsequent planetary accretion \citep{helled_giant_2014}. \cite{molliere_interpreting_2022} list the numerous pitfalls and challenges in trying to invert the formation history of individual planets based on present-day atmospheres. Therefore, extreme caution must be exercised in attempts to connect the present-day atmospheric observations with formation and evolution history. Instead it is possible that planetary mass -- atmospheric (and bulk) metallicity trends across a homogeneous sample of planets, sampled across planetary mass and stellar mass space, present a better alternative \citep{thorngren_mass-metallicity_2016, teske_metal-rich_2019, welbanks_mass-metallicity_2019}. 

We can now begin to test the limits of these current theories by atmospheric characterization of carefully selected samples across a larger stellar mass range with JWST and ARIEL in the future \citep{tinetti_science_2016}.

\section{\textit{Searching for GEMS} Survey Outline}\label{sec:outline}
Considering the theoretical considerations (and limitations) presented above, and the observational tests to discriminate between the formation of GEMS in the protoplanetary (through core-accretion) vs.~protostellar (through gravitational instability), we ask the following questions:
\begin{enumerate}
    \item \textit{Mass-Radius+ (M-R+) relations:} Does the difficulty in forming GEMS in this mass starved regime through core-accretion manifest as a systematic difference in their bulk-properties as a function of stellar mass, i.e., characterizing the sample of giant planets in 3D (planet mass, planet radius, stellar mass)?
    \item \textit{Occurrence Rates:} Do GEMS have a lower occurrence compared to similar giant planets in short orbital periods (the canonical hot Jupiters) around FGK stars? In particular, the M-dwarf spectral type also allows comparison of the giant planet occurrence rate as a function of spectral sub-type.
\end{enumerate}

To address these questions, we have started the \textit{Searching for GEMS} survey, which is currently ongoing, and entails the follow-up and confirmation of new GEMS planet candidates in addition to performing demographical analysis on TESS light-curves to provide improved constraints on the occurrence of short-period GEMS. This transit survey will also contribute GEMS to answer the question: \textit{Does the metallicity transition in the planetary mass -- stellar metallicity plane between core-accretion and GI (Section \ref{sec:stellarmetallicity}) also depend on the host-stellar mass?} To answer this we will extend the 2D planetary mass -- stellar metallicity plane to a third dimension to include stellar mass.

\subsection{Requirements for M-R+ relations}\label{sec:required_sampleMR}
In order to answer the question posed for M-R+ relations we generate a set of simulated catalogues of GEMS with mass measurements based on different assumptions, and assess our ability to recover the input trends as a function of catalogue size.

\begin{figure}
\centering
\includegraphics[width=\columnwidth]{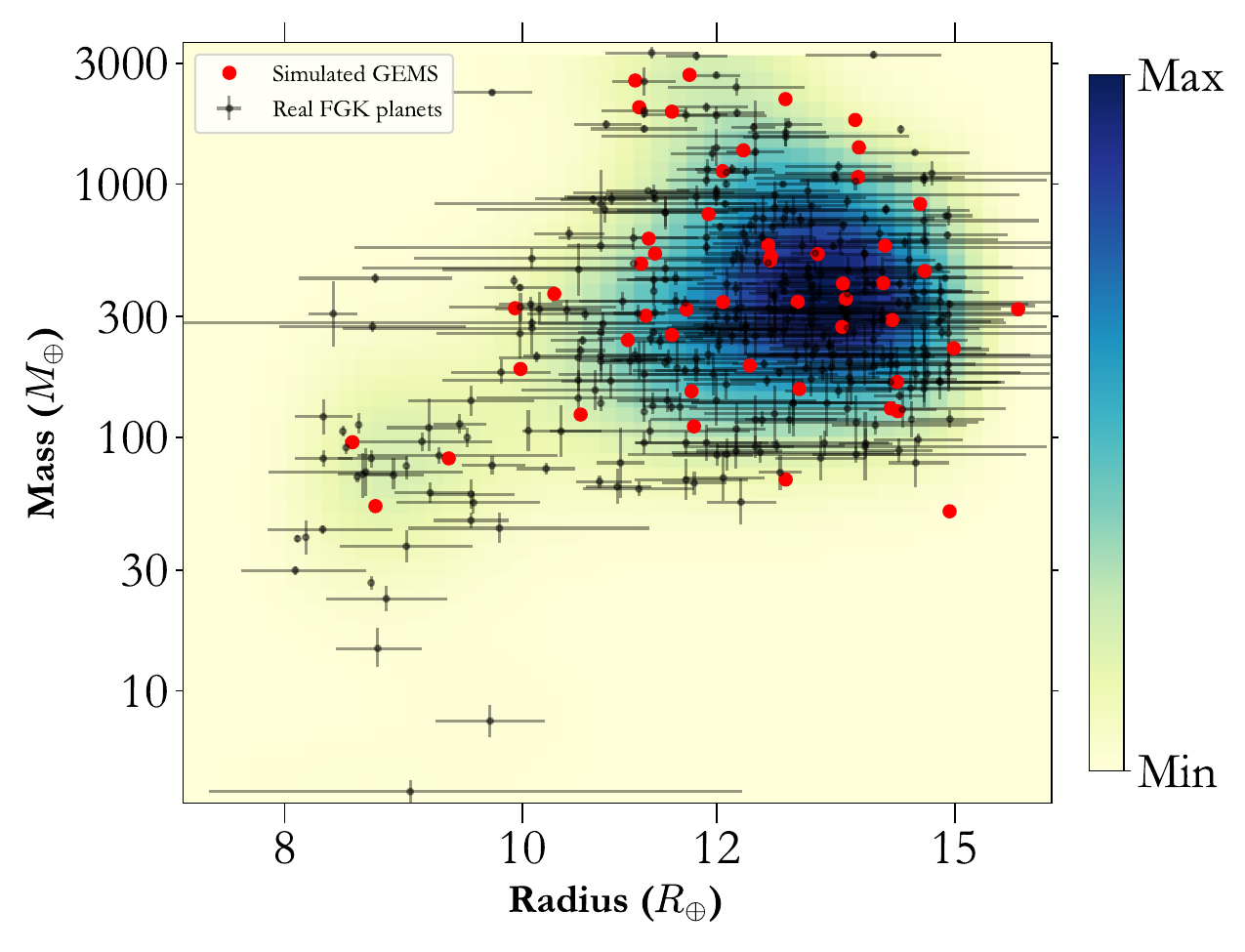}
\caption{The 2D joint probability distribution --- \textit{f(m, r)}$_{FGK}$ ---  for $\sim$ 350 giant planets orbiting FGK stars shown in the background, where the black points are the real measurements of FGK planets, the red dots are an example simulation of 50 GEMS obtained via rejection sampling of the underlying FGK PDF, and the underlying colour is their probability. Each of these simulated GEMS sample of planets is compared with the FGK sample to estimate the minimum sample size required for the survey as described in Section \ref{sec:2dmrsim}.}
\label{fig:2DJointMR_FGK}
\end{figure}

\begin{figure*}[ht]
\fig{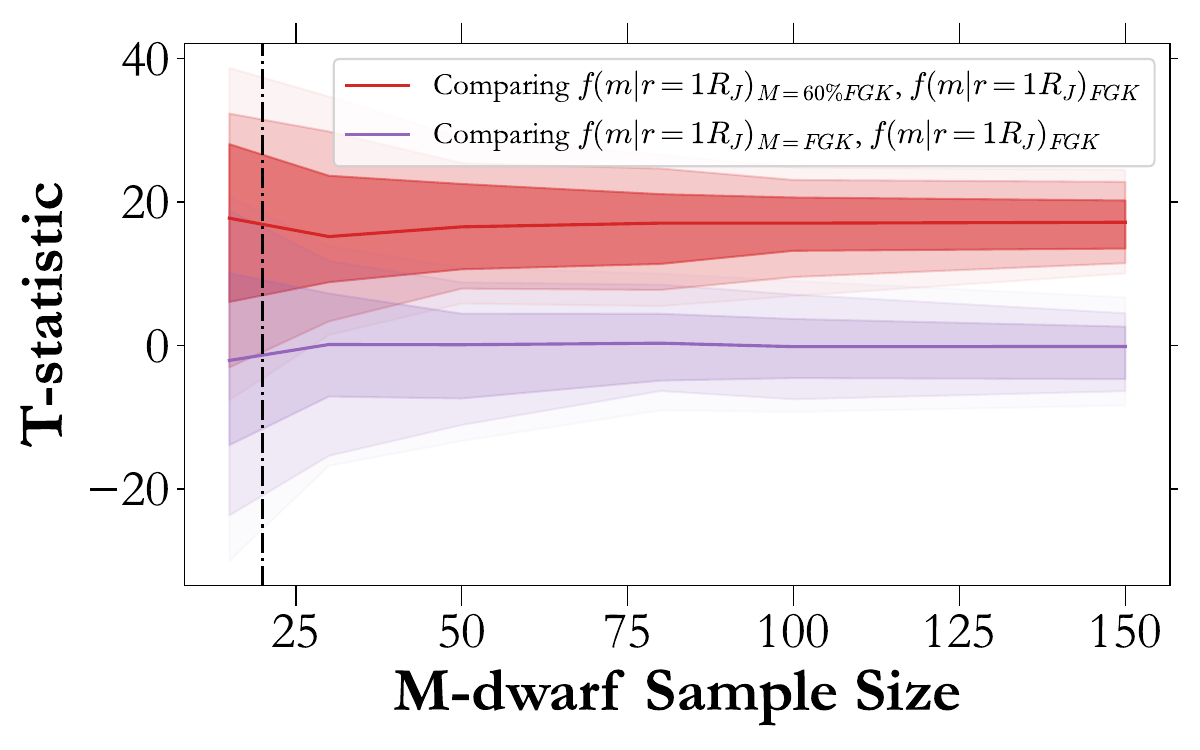}{0.45\textwidth}{}
\fig{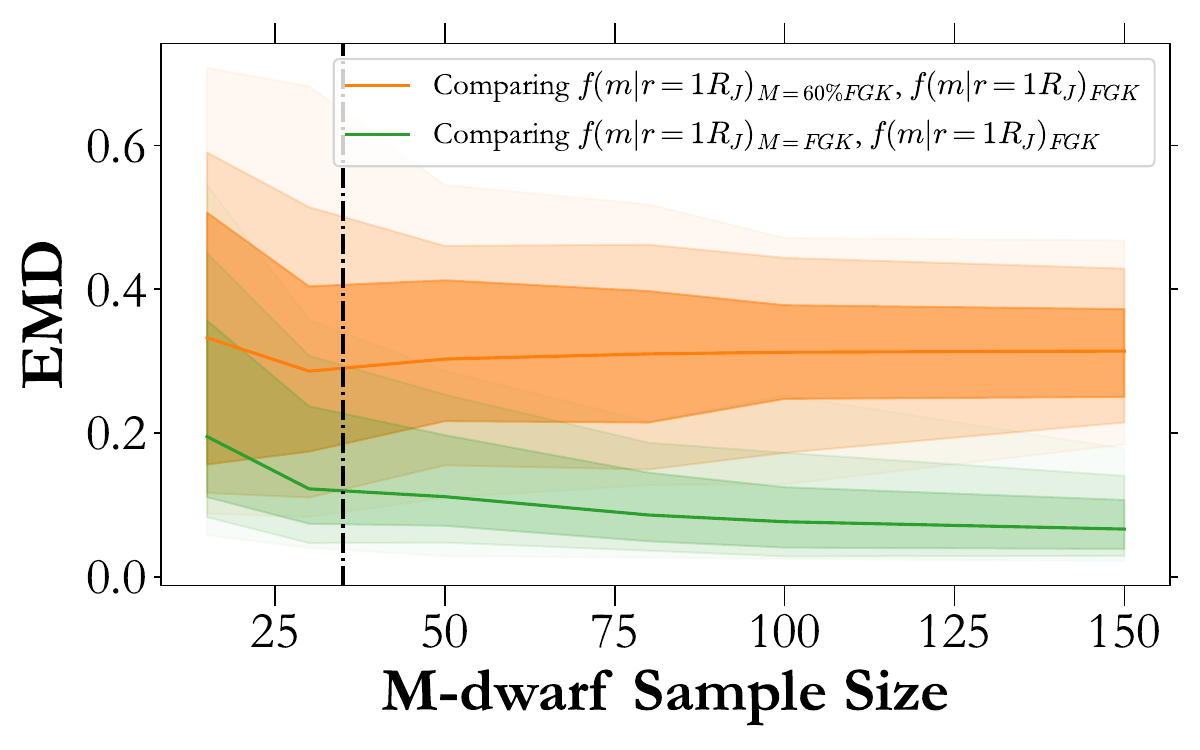}{0.45\textwidth}{}
\caption{\small The solid line represents the median metric value, while the different shaded regions represent the 16-84\%, 5-95\%, and 1-99\% percentile regions respectively. The vertical black line is the minimum sample size of transiting GEMS required based on the comparing the distributions. \textbf{Left:}  Welch's t-test suggesting the need for about 20 GEMS if we solely wish to compare the median values of the two distributions assuming normality.  \textbf{Right:} The Earth-mover distance, which suggests the need for about 35 GEMS to distinguish between the two distributions agnostic of Gaussian assumptions. \textbf{Takeaway:} Identifying the minimum sample size required to distinguish between a sample of giant planets around FGK stars and GEMS based on their M-R distribution when the GEMS are about 60\% in mass of the FGK planets, we require about 35 GEMS if we do not assume an underlying normal distribution and 20 GEMS if we do.}\label{fig:2dSimulation}
\end{figure*}

\subsubsection{Mass-Radius (M-R)}\label{sec:2dmrsim}
We start off by trying to recover and quantify differences in the 2D mass-radius (M-R) distribution of giant planets between a sample of FGK stars (
$\sim$ 0.6 \solmass~-- 1.5 \solmass) and a simulated sample of GEMS ($\lesssim$ 0.6 \solmass). Our FGK sample consists of $\sim 350$ transiting giant planets (15 \earthmass{} $\gtrsim$ $R_p$ $\gtrsim$ 8 \earthmass{}) orbiting stars ranging from $\sim$ 0.6 \solmass{} to 1.5 \solmass{} with planetary masses known to better than 3-$\sigma$ \citep{akeson_nasa_2013, PSCompPars}. The steps followed are as follows:

\begin{enumerate}
    \item \textbf{Obtain joint mass-radius distribution --- \textit{f(m, r)}$_{FGK}$ --- for FGK giant planet sample:}  We fit the FGK sample using the updated non-parametric inference tool \texttt{MRExo} \citep{kanodia_mass-radius_2019, kanodia_beyond_2023}, which utilizes beta-density (or normalized Bernstein) polynomials \citep{ning_predicting_2018} to jointly fit their masses and radii and obtain the 2D probability density function (PDF) --- \textit{f(m, r)} --- where \textit{m, r} signify the planetary mass and radius respectively. We use the cross-validation functionality within this toolkit to optimize for the number of degrees (for the beta density functions) to be 48 in each dimension \citep{kanodia_mass-radius_2019, kanodia_beyond_2023}.

    \item \textbf{Simulated M-dwarf sample:} We perform rejection sampling in 2D to obtain a simulated M-dwarf sample with \textit{n} planets in mass-radius space following the same distribution as the FGK sample, to which we ascribe 10-$\sigma$ and 5-$\sigma$ errors in radius and mass respectively. We then use \texttt{MRExo} to fit the joint M-R distribution --- \textit{f(m, r)}$_{M=FGK}$ --- for this simulated sample. We repeat this step 100x times to obtain a distribution of PDFs for the simulated M-dwarf catalogue across 6 steps spanning a range of GEMS sample sizes from 15 to 150 (\autoref{fig:2DJointMR_FGK}). We simulate the entire M-dwarf catalogue from scratch (instead of training it on the existing transiting GEMS sample) to avoid any potential biases that could be incurred from this limited sample size, and estimate the ideal sample-size in an agnostic manner. 
    
    \item \textbf{Obtain the conditional distribution of planetary mass for a given radius --- $f(m | r)$ --- for real FGK and simulated M-dwarf samples:} The 2D joint distribution --- \textit{f(m, r)} ---  can then be conditioned on Jupiter's radius ($R_J$ or 11.2 \earthradius) to obtain a PDF for the inferred mass of Jovian-sized objects --- $f(m | r=1 R_J)$. The ability to compare $f(m | r=1 R_J)_{FGK}$ and $f(m | r=1 R_J)_{M=FGK}$ in this simulated case where we have the same joint M-R distribution for the two (i.e., the conditional distributions should be similar), represents our `best-case scenario' in being able to distinguish between two the inferred PDFs for two samples of a given size.
    
    \item \textbf{Metrics to compare the two 1-D PDFs:} We then perform rejection sampling in 1-D of the normalized PDFs --- $f(m | r=1 R_J)$ --- to obtain a histogram of 1000 simulated planetary masses to compare across the M-dwarf and FGK samples. We adopt the Earth-mover distance \citep[EMD;][]{rubner_metric_1998, ramdas_wasserstein_2015}, which is a metric to quantify the difference between two distributions without any assumptions of normality or similar variance. In other words, the EMD can be used to compare two distributions, and reduces to zero when they are completely identical. The EMD is identical to the Wasserstein distance \citep{kantorovich_mathematical_1960, ramdas_wasserstein_2015} in the case of PDFs and is implemented in the \texttt{scipy python} package \citep{oliphant_python_2007, virtanen_scipy_2020}. We do not adopt the Kullback-Leibler (KL) divergence since that assumes comparison of a sample with a known distribution \citep{kullback_information_1951}, and is asymmetric. For the purposes of comparing just the mean of the two PDFs (while assuming normality), we adopt Welch's t-test \citep{welch_generalization_1947, ruxton_unequal_2006} implemented in \texttt{scipy} through the \texttt{ttest\_ind} function that allows for unequal variances. This is in lieu of the two-sample Kolmogorov-Smirnov \citep[K-S; ][]{kolmogorov_sulla_1933, smirnov_table_1948} and Anderson-Darling \citep[AD; ][]{anderson_asymptotic_1952, scholz_k-sample_1987} tests.
    
    \item We then repeat steps 2 and 3, where we reduce the masses for the M-dwarf planet sample to 60\% and 80\% of the FGK sample to quantify the ability to distinguish between the M-dwarf and FGK samples as a function of M-dwarf sample size.
    
    \item Finally we calculate the EMD and t-statistic for comparing histograms of the M-dwarf and FGK samples, and pick the optimum sample size where the 5th percentile of our metric comparing our test distribution with the FGK sample ($f(m | r=1 R_J)_{M = 80\% FGK}$, $f(m | r=1 R_J)_{FGK}$) intersects with the 50th percentile (median) of the metric distribution from our best-case scenario with similar distributions ($f(m | r=1 R_J)_{M =FGK}$, $f(m | r=1 R_J)_{FGK}$).
\end{enumerate}

Based on the criterion defined above, we are able to distinguish between mean of the M-dwarf and FGK planetary M-R distributions when the M-dwarf distribution is reduced to 80\% and 60\% (i.e., a 20\% and 40\% difference) with about 100 and 20 GEMS respectively using Welch's t-test. Using the more agnostic and model-independent EMD, we require 35--40 GEMS to distinguish between the two distributions when the M-dwarf masses are offset to 60\%, and > 150 for the 80\% case (\autoref{fig:2dSimulation}). Thus limiting the comparison between the two samples to just the M-R plane would require about 35 GEMS with mass measurements at the 5-$\sigma$ level to ascertain a 40\% difference in mass.

\subsubsection{Mass-Radius-Stellar Mass (MRStM)}
\begin{figure}[!b]
\centering
\includegraphics[width=\columnwidth]{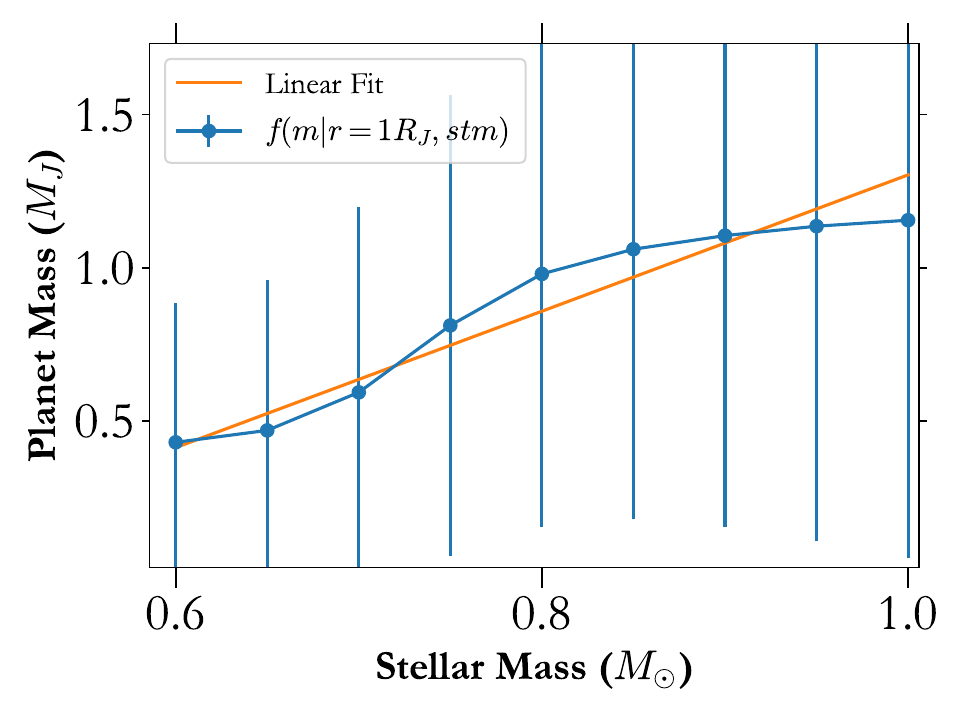}
\caption{The scatter plot shows the expectation value --- \textit{f(m| r, stm)} --- for the Jovian sized objects orbiting different stellar masses, whereas the orange line shows the linear fit, which we then extrapolate to lower stellar masses for M-dwarfs.}
\label{fig:FGK_StMass_PlMass}
\end{figure}

Instead of restricting our analysis to just the two M-R dimensions, we can incorporate additional information about these systems in the form of their stellar mass to jointly fit \textit{f(m, r, stm)}, where \textit{stm} refers to the continuous variable spanning stellar mass using \texttt{MRExo}. Using the sample defined above, we perform a fit in 3D where the cross-validation method optimizes for the number of degrees to be 41 in each dimension.   Conditioning on planetary radius and stellar mass to predict the planetary mass --- $f(m| r=1 R_J, stm)$ --- across a range of stellar masses, we obtain a nominal linear trend between the expectation values of the PDFs and stellar mass as shown in \autoref{fig:FGK_StMass_PlMass}.

\begin{figure*}[ht]
\fig{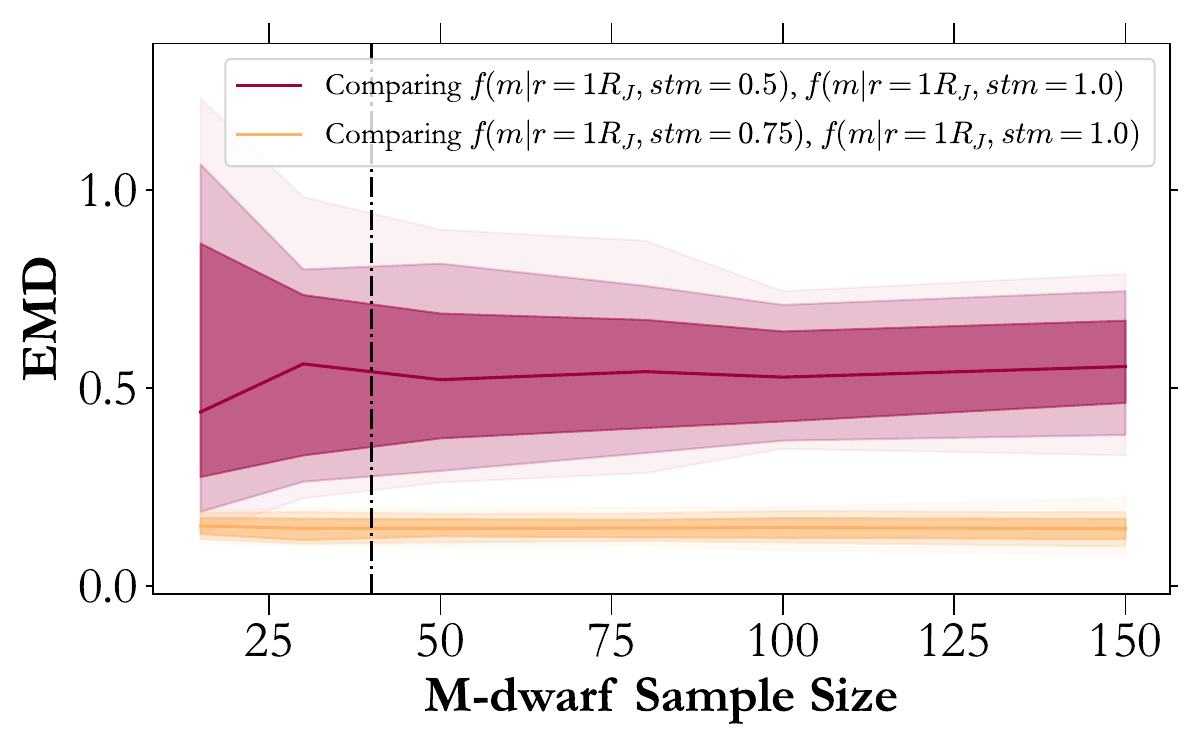}{0.5\textwidth}{}
\fig{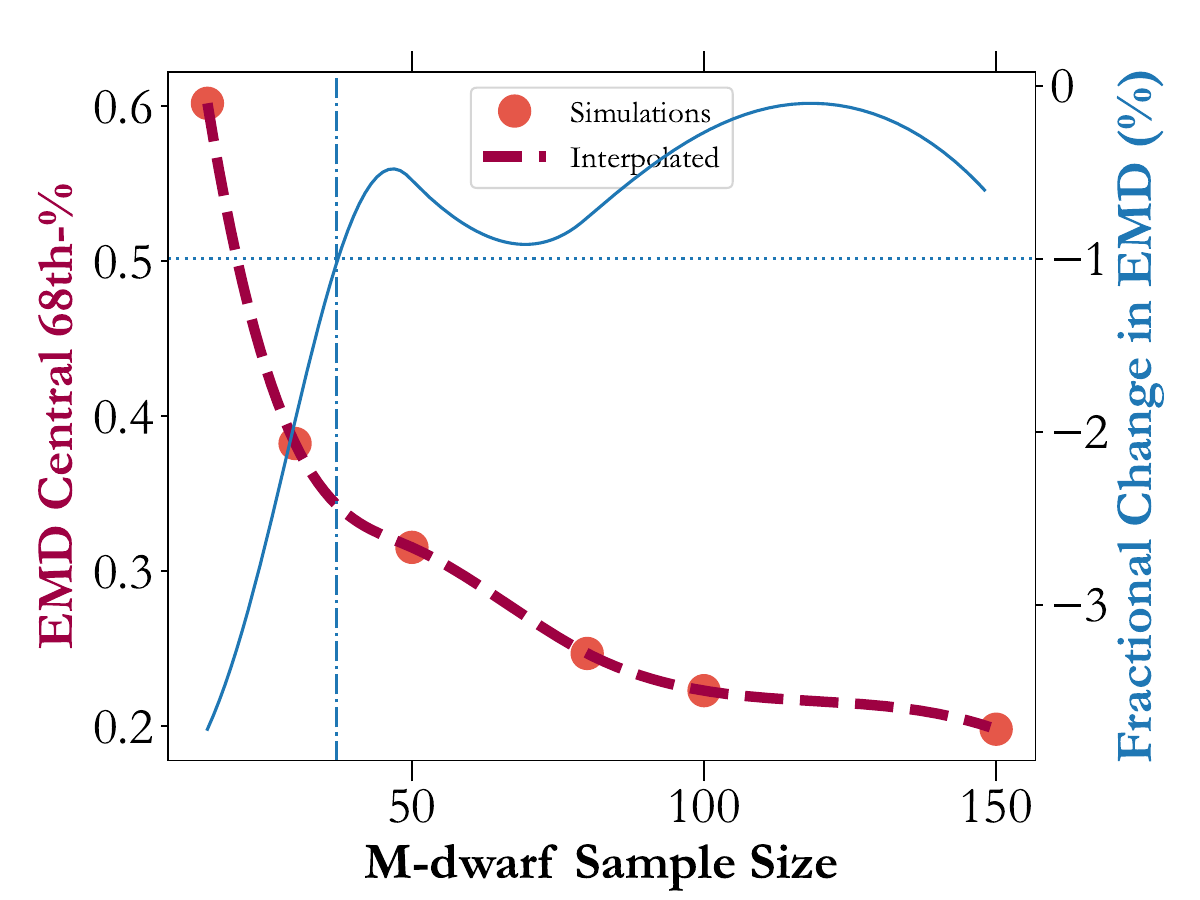}{0.45\textwidth}{}
\caption{\small \textbf{Left:} Similar to \autoref{fig:2dSimulation}, we plot the EMD while comparing $f(m,r|stm = 0.75~$\solmass) with $f(m,r|stm = 1.0~$\solmass) in orange, and then $f(m,r|stm = 0.5~$\solmass) with $f(m,r|stm = 1.0~$\solmass) in red. Since the K-dwarf sample is fixed in sample-size, it does not show any dependence on M-dwarf sample-size, whereas the red region converges. \textbf{Right:} To establish convergence of the EMD as a function of sample-size (i.e., minimum sample-size beyond which the improvement in EMD is marginal), we consider the distance between the central 68-th percentile of the EMD distribution as a function of sample-size and the sample-size where its fractional change goes below 1\%. \textbf{Takeaway:} The EMD metric to ascertain the stellar dependence for  $f(m| r=1 R_J, stm)$ starts to converge for a GEMS sample of $\sim$ 40.}\label{fig:3dSimulation}
\end{figure*}

Similar to the methodology from the 2D simulation, we generate synthetic M-dwarf GEMS samples and compare the predicted masses for these planets with those around more massive stars using the EMD metric. We do not investigate the uncertainties associated with the expectation values here (\autoref{fig:FGK_StMass_PlMass}), since the significance of this trend (both statistical and physical) will be discussed in a follow-up paper. We first generate a synthetic M-dwarf catalogue where the 2D joint $f(m,r)$ distribution is same as that for a solar mass star, i.e., $f(m,r|stm = 1.0~$\solmass). For this synthetic catalogue of planets, we ascribe stellar masses using a half-normal distribution $\equiv$ 0.6 - $|\mathcal{N}$(0, 0.1)$|$, to avoid an undue bias towards mid-to-late M-dwarfs or include K-dwarfs. Then the mass for each M-dwarf planet is scaled based on an extrapolation of the linear trend shown in \autoref{fig:FGK_StMass_PlMass}, which is further normalized to be $\sim$ 1 for Jupiters around solar-mass stars. This way we ascribe a stellar mass dependence to the simulated GEMS, and generate 100 such catalogues for each M-dwarf sample size ranging from 15 to 150 and append these to the original FGK sample. This dependence is partly motivated by the existing sample, and will be explored in further detail (both empirically and theoretically) in part II of this work. After performing a 3D fit --- \textit{f(m, r, stm)} --- for each catalogue of planets, we compare the $f(m,r|stm = 0.75~$\solmass) with $f(m,r|stm = 1.0~$\solmass) as a control that should not show any sample-size dependence (since fixed FGK sample), and then $f(m,r|stm = 0.5~$\solmass) with $f(m,r|stm = 1.0~$\solmass) to show the convergence of the EMD with increasing the M-dwarf sample size (\autoref{fig:3dSimulation}).

To establish convergence, we consider the M-dwarf sample size where the fractional change of the central 68th percentile (EMD$_{84\%}$ - EMD$_{16\%}$) of the EMD distribution with sample-size goes below 1\%, which is at $\sim 40$. To test the impact of the assumed slope on the inferred sample-size, we also compare the EMD for two $f(m| r=1 R_J)$ distributions for stellar masses of 0.55 \solmass{} and 0.6 \solmass{}, i.e., with more similar distributions, and find that we attain a similar 1\% convergence criterion with a GEMS sample of about 40 -- 50. 

These simulations help us assert that we should be able to distinguish between different stellar-mass -- planetary mass dependencies for Jovians with a sample size of about 40 -- 50 planets, which will help answer the first question regarding M-R+ relations --- whether the difficulty in forming GEMS through core-accretion in a mass starved environment manifests as a systematic difference in their bulk-density as a function of stellar mass.


\subsection{Limitations of Existing Transiting GEMS Occurrence Estimates}
\cite{gan_occurrence_2023} have estimated the occurrence of short-period (0.8 -- 10 days) transiting giant planets (7 \earthradius $\le R_p \le$ 2 $R_J$) around $\sim$ 60,000 early M-dwarfs with \teff~ between 2900 K and 4000 K, and stellar masses between 0.45 \solmass~and 0.65 \solmass and $10.5 \le T_{mag} \le 13.5$. as 0.27 $\pm$ 0.09 \%. Similarly, \cite{bryant_occurrence_2023} suggest an occurrence rate of 0.194 $\pm$ 0.072 \% for planets with (0.6 $R_J$ $\le R_p \le$ 2.0 $R_J$) orbiting $\sim$ 90,000 nearby low-mass stars with $M_* \le$ 0.71 \solmass. These investigations have already started to provide insight into the occurrence of transiting GEMS, however we note three main limitations with these existing studies that our survey design aims to improve upon:

\begin{enumerate}
    \item Jovian-sized objects can range in mass from 0.3 $M_J$ to $\sim$ 100 $M_J$, i.e, Saturn-massed to late M-dwarfs. It is not possible to confirm the planetary nature of transiting Jovian-sized objects, with statistical validation alone (i.e., without dynamical confirmation with RVs or TTVs). Therefore, the transiting objects used in these two studies to estimate occurrence rates are not necessarily all planets and likely contaminated by astrophysical false-positives such as brown-dwarfs and eclipsing binaries (EBs). This is indeed seen for TOI-5375~B that was identified as a planet candidate by \cite{gan_occurrence_2023}, but confirmed to be an EB by \cite{lambert_toi-5375_2023}. These studies have attempted to account for this by assigning a false positive probability (FPP) for the candidates discovered through their pipelines using statistical validation tools \citep{bryant_occurrence_2023} or estimating this based on the prevalence of brown-dwarfs and EBs in the literature \citep{gan_occurrence_2023}. However, literature samples of brown-dwarfs and giant planets are heterogeneous and not drawn from M-dwarf hosts, additionally, as pointed out by \cite{bryant_occurrence_2023}, the statistical techniques are limited in their ability to distinguish between giant planets, brown-dwarfs and very low-mass stars. Without more relevant FPP estimates (from M-dwarf host, short orbital period and homogeneous samples), or spectroscopic validation, these existing estimates should be considered carefully (and with the appropriate caveats) while comparing across samples and surveys. 
    \item The input sample of M-dwarfs for both studies conflates mid-to-late K-dwarfs with early M-dwarfs. For example, \cite{bryant_occurrence_2023} have a \teff{} and stellar mass upper limit of 4500 K and 0.75 \solradius~respectively, which includes mid-to-late K-dwarfs. Given the potentially positive correlation between stellar-mass and the occurrence of GEMS \citep{johnson_giant_2010, zhou_two_2019, bryant_occurrence_2023}, this choice for the input stellar sample biases the occurrence rate to higher values. \cite{bryant_occurrence_2023} account for this by dividing their stellar sample into different stellar mass bins.
    \item  These studies utilize a sample of $<$~100,000 M-dwarfs analyzing light-curves for bright(-er)  M-dwarfs as reduced by the Science Processing Operations Center \citep[SPOC; ][]{jenkins_tess_2016, caldwell_tess_2020} or Quick Look Pipeline \citep[QLP;][]{huang_photometry_2020, kunimoto_qlp_2021} efforts. In particular, the \cite{gan_occurrence_2023} sample is (mostly) magnitude limited, but confounded by the selection function of the QLP sample, and consists of 60,000 (primarily) early M-dwarfs. Conversely, the \cite{bryant_occurrence_2023} sample is volume limited to 100 pc, but also includes cuts based on magnitude, which precludes it from being volume complete across the entire M-dwarf spectral type\footnote{For example, TRAPPIST-1 at 100 pc has a $T$ magnitude of $\sim$ 18 mag, which is fainter than their magnitude cut of 16 mag.}. Furthermore, the distance and magnitude cut-offs chosen for the above studies do not include most of the confirmed planetary transiting GEMS that have been discovered thus far and are beyond 100 pc (\autoref{tab:InputSample}).  
\end{enumerate}

Despite these limitations, these existing occurrence estimates already demonstrate the lower occurrence of transiting GEMS compared to FGK hot-Jupiters. We will attempt to tackle these limitations through our systematic candidate search and analysis that we describe in the survey work plan.

\subsection{Survey Design}\label{sec:cuts}

While the methodology and preliminary results from this ongoing survey will be described in upcoming manuscripts, we briefly describe the sample selection, work plan and survey status here.

\subsubsection{Sample Selection}

\begin{figure*}[ht]
\centering
\fig{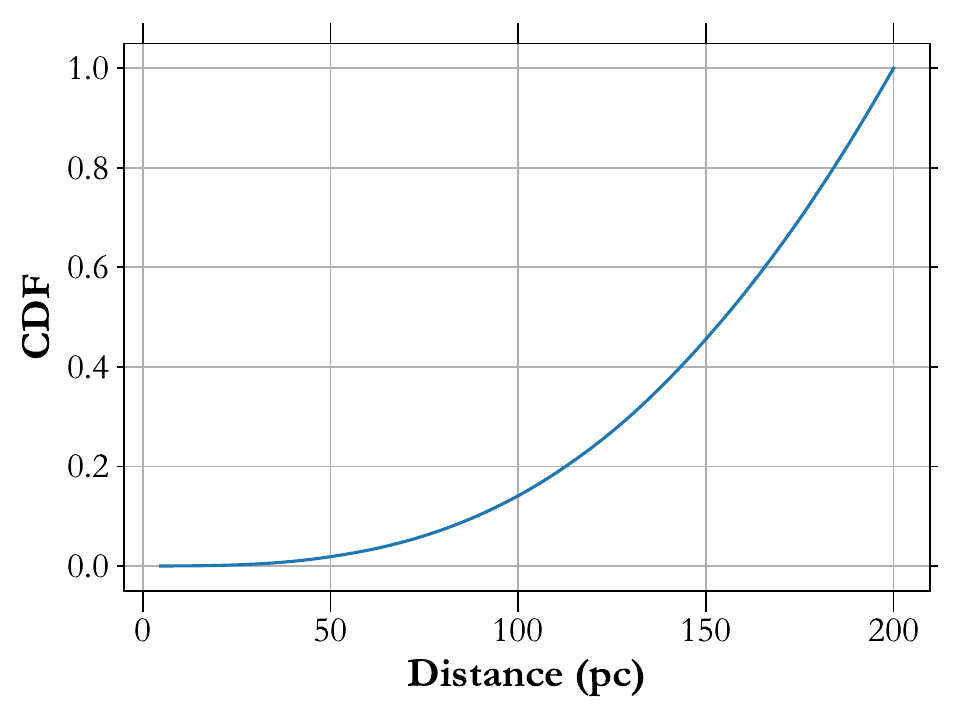}{0.45\textwidth}{\small (a) Cumulative distribution function (CDF) of the distance.}
\fig{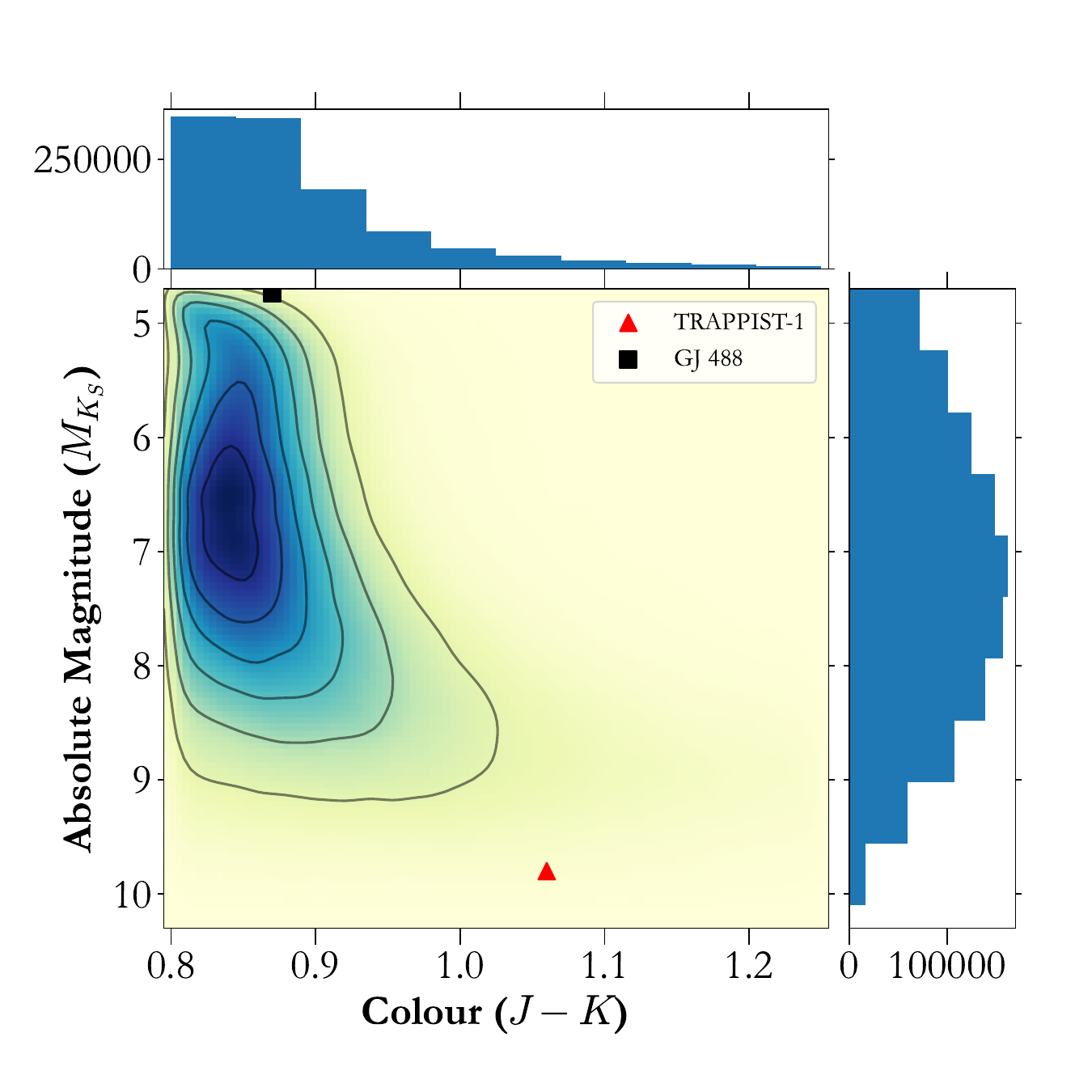}{0.45\textwidth}{\small (b) Color - absolute magnitude diagram.}
\fig{Histogram_Rstar.pdf}{0.3\textwidth}{\small (c) Histogram of Stellar Radius.}
\fig{Histogram_Mstar.pdf}{0.3\textwidth}{\small (d) Histogram of Stellar Mass.}
\fig{Histogram_Teff.pdf}{0.3\textwidth}{\small (e) Histogram of Effective Temperature.}
\fig{CDF_Jmag.pdf}{0.3\textwidth}{\small (f) CDF of J magnitude.}
\fig{CDF_Kmag.pdf}{0.3\textwidth}{\small (g) CDF of K magnitude.}
\fig{CDF_ContRatio2.pdf}{0.3\textwidth}{\small (h) CDF of TIC contamination ratio.}
\caption{\small The $\sim1.0$ million stars observed by TESS through Cycle 5 and contained in our cuts in Section \ref{sec:cuts}. Previous studies have used an upper limit of 100 pc, which represents a small fraction ($< 20$\%) of our sample. Our sample spans the full range of color and magnitude for M0-M9 . For reference, we indicate GJ 488 (M0V; square) and TRAPPIST-1 (M8V; triangle) in the CMD to approximately denote the two ends of the M-dwarf spectral type. The stellar radii, masses and \teff{} are calculated from $M_{K_s}$ using relations from \cite{mann_how_2015, mann_how_2019}, whereas the TICv8.2 contamination ratio is from \cite{stassun_tess_2018}. }\label{fig:SampleCut}
\end{figure*}

We adopt a volume limited sample spanning 200 pc, which covers most of the existing confirmed planets. Since confirming the planetary nature of transiting Jovian objects requires resource-intensive observations, this approach capitalizes on existing observations while providing a large sample for more precise occurrence estimates. This is to help overcome the limitations of existing efforts described above, primary among which is the need for spectroscopic validation of the planet candidates utilized in occurrence rate estimates, in the absence of which, the need for more informed and relevant false positive estimates. To do this, we start off with the Gaia DR3 catalogue \citep{gaia_collaboration_gaia_2023}, to which we apply the following cuts:
\begin{enumerate*}[label=(\roman*)]
\item a parallax of $\varpi>5\mathrm{~mas}$ (corresponding to an upper limit of 200 pc),
\item a parallax error of $\sigma_\varpi<1\mathrm{~mas}$, and
\item null quality flags indicating the object is not a quasi-stellar object or galaxy (not in the Gaia DR3 \texttt{qso\_candidates} or \texttt{galaxy\_candidates} tables), not a binary star (not identified in Gaia DR3 as an astrometric, spectroscopic, or EB), and not a duplicated source. 
\end{enumerate*} 

The source identifier in each Gaia data release is not guaranteed to be the same and we obtain the Gaia DR2 source identifier using the \texttt{dr2\_neighbourhood} table provided with the Gaia DR3 release. We reject sources without matching \texttt{dr2\_source\_id} and \texttt{dr3\_source\_id} because this can arise from a spurious detection in either DR2 or DR3 \citep{Torra2021} or erroneous proper motions from DR2 and double stars that are a resolved pair in Gaia DR3 \citep{EDR3Valid}.
We cross-match the Gaia objects with \texttt{TICv82} using the Gaia DR2 source identifier \citep{stassun_revised_2019} and retain stars observed through TESS Cycle 5 using the high-precision TESS pointing tool\footnote{\url{https://github.com/tessgi/tess-point}} \citep{tesspoint}. To select the M dwarfs in the sample, we place constraints on the 2MASS colors and magnitudes \citep[Table 7;][]{cifuentes_carmenes_2020} keeping stars with \begin{enumerate*}[label=(\roman*)]
\item $0.8<J-K<1.25$ and
\item $4.7<M_{K_s}<10.1$
\end{enumerate*}, resulting in our final sample of $\sim 1$ million stars, the properties of which are depicted in \autoref{fig:SampleCut}. Splitting this along the spectral sub-type using $M_{K_s}$ cut-offs from \cite{cifuentes_carmenes_2020}, we note that about 21\% of our sample are early M-dwarfs (< M2.5V; $M_{K_s}$ < 6), 30\% as mid M-dwarfs (M2.5V -- M4; 7.1 > $M_{K_s}$ > 6), with the rest being late M-dwarfs. While we denote GJ 488 and TRAPPIST-1 as example M0/M9s (\autoref{fig:SampleCut}), these do not set our CMD cut-offs due to the intrinsic astrophysical scatter ($\sim$ 0.2 mag) in the colour of early M-dwarfs as seen in  Figure 14 and A2 from \cite{cifuentes_carmenes_2020}. We use the CMD to define our sample instead of derived stellar parameters which often suffer from inaccuracies due to systematics in internal structure models and evolutionary tracks \citep{kesseli_magnetic_2018, dieterich_solar_2021, passegger_metallicities_2022}.

\subsubsection{Work Plan}

Given their faintness, publicly available light curves such as from SPOC or QLP do not exist for our entire sample. Instead we will extract light curves from the full-frame image (FFI) images using the public package TESS-Gaia Light Curves \citep[\texttt{TGLC}; ][]{han_tess-gaia_2023} for all our targets, which performs a contamination correction for the TESS light curves based on their Gaia positions, magnitude and colour. A detailed characterization of the photometric performance of this reduction routine for our sample as a function of stellar properties, will be included in a future manuscript upon completion of said search.Then, we will search for transiting GEMS candidates around our M-dwarf sample using the box-least squares algorithm \citep{kovacs_box-fitting_2002} to identify candidates and a combination of publicly available tools to vet these candidates for astrophysical false positives \citep[e.g., \texttt{DAVE} and \texttt{TRICERATOPS};][]{kostov_l_2019,giacalone_vetting_2021}. The candidates that survive vetting will be observed from ground-based facilities to obtain transits (to rule out background EBs), high-contrast imaging (to constrain dilution) and spectra (to rule out EBs and brown-dwarfs). This sample of well-vetted planets will be used to inform the occurrence rates for GEMS.

To estimate the number of true GEMS we will obtain here, we have assumed an occurrence rate of 0.1~--~0.2\% alongside a transit probability of 5 -- 10\%; the occurrence rate is from a 0.4 \% estimate for AFG stars from \cite{zhou_two_2019}, which is then scaled by a stellar mass dependent ratio. Additionally, this estimate also agrees with recent upper limits from \cite{gan_occurrence_2023} and \cite{bryant_occurrence_2023}. We note that since we already have $\sim 15$ confirmed transiting GEMS (\autoref{tab:InputSample}), we can already place a lower-limit on the occurrence of short-period transiting GEMS at $\sim 0.03$\% (after accounting for transit probability). While we do not have good priors on the astrophysical false-positives (brown-dwarfs, EBs) around M-dwarfs at these orbital periods, we assume a 1:1 ratio between bonafide planets and false positives here.

Based on these estimates, we expect to discover about $\sim$ 100 transiting GEMS, the spectroscopically validated sub-sample (roughly half) of which will enable accurate (due to spectroscopic mass upper limits) and precise (due to the $>$ 10x larger input sample-size) occurrences. The large sample-size will also enable estimating the occurrence rate as a function of stellar mass across the M-dwarf spectral sub-type. To convert our detection efficiency into occurrence rates, we will follow previous studies by performing injection \& recovery tests to quantify the search sensitivity \citep{dressing_occurrence_2015, gan_occurrence_2023, bryant_occurrence_2023, ment_occurrence_2023}. The parameters, methodology, and results from these tests will be published along with the final occurrence rates.

Despite using \texttt{TGLC} which performs a dilution correction based on Gaia magnitudes, to estimate the yield from our sample we assume a TICv8.2 contamination ratio \citep{stassun_tess_2018} cut-off\footnote{Where the flux from contaminating background objects is equal to the flux from the target object.} of 1.0, beyond which the field is likely to be too crowded for follow-up.

The bulk of our spectroscopic follow-up will be performed with the Habitable-zone Planet Finder \citep[HPF;][]{mahadevan_habitable-zone_2012, mahadevan_habitable-zone_2014} on the 10-m Hobby Eberly Telescope \citep{ramsey_early_1998}, NEID on the 3.5-m WIYN telescope \citep{halverson_comprehensive_2016, schwab_design_2016}, alongside MAROON-X on the 8-m Gemini-N \citep{seifahrt_maroon-x_2022}, and the Planet Finding Spectrograph (PFS) on the 6.5-m Magellan Clay telescope \citep{crane_carnegie_2006, crane_carnegie_2008, crane_carnegie_2010}. Based on the existing observations and the exposure time calculator for HPF\footnote{\url{https://psuastro.github.io/HPF/Exposure-Times/}},  we estimate being able to perform spectroscopic validation for faint M-dwarfs going down to $J$ $< 15$, which combined with the declination limits for the telescope of approximately -10$^{\circ}$ to 72$^{\circ}$, and the contamination ratio cut-off defined above, covers about half our sample of a million M-dwarfs. In addition to this, we expect to be able to observe an additional fraction of stars with NEID, MAROON-X, which despite being optical/red-optical instruments operate on conventional telescopes with more lenient declination limits. Similarly PFS operates in the Southern hemisphere, but uses the iodine gas-cell technique for RV determination \citep{butler_attaining_1996}, making it less suitable for following-up faint, red M-dwarfs. Given the vagaries of telescope time allocation, we conservatively assume that using a combination of HPF, NEID, MAROON-X and PFS, at least half the stars in our 200 pc sample will be amenable to spectroscopic validation. Therefore, any planet candidates discovered in this half of the sample should be available for validation and potential confirmation. The faintness limit of HPF allows us to be almost volume limited (covers $>$ 90\% of the sample) in the region of the sky accessible to HET, and derive empirical FPP for our sample, which we can then apply to the rest of the candidates that might be too faint for follow-up from the other facilities.

A subset of the validated planets ($\sim 40$ as motivated in previous sections, of which 15 have already been confirmed) will be selected for more intensive follow-up to obtain planetary mass measurements, which can then be utilized for future investigations that will be discussed in our follow-up paper.  Through ongoing observations of objects classified as TOIs or CTOIs that also form part of our 200 pc sample, the community has confirmed $\sim 15$ GEMS so far, in addition to which we have also identified over 20 astrophysical false positives that will be discussed in a follow-up paper discussing preliminary results and trends from our survey.

\subsubsection{Survey Status}
The ongoing survey has currently led to the discovery and confirmation of nine of the transiting GEMS indicated in \autoref{tab:InputSample}, with five additional planet confirmation manuscripts currently in preparation. The demographics analysis has been initiated, with the sample selection (as explained above) complete, and vetting procedure being finalized and trained based on a test run on a 100 pc sub-sample. Results from the 100 pc sample will be published as an intermediate paper, which will also serve as a comparison with previous works such as those by \cite{bryant_occurrence_2023}. Candidates from these searches are simultaneously being validated and followed-up from ground-based telescopes to confirm their planetary nature, and estimate their stellar and planetary properties. After finalizing the candidate search and vetting procedure, we will also perform injection and recovery tests to quantify our detection sensitivity, and go from candidate detection, to measured occurrences. Upon completion of the survey and confirmation of planet candidates, we will perform the statistical analysis outlined in previous sections to quantify the properties of the observed transiting GEMS and compare them with similar transiting giant planets around more massive stars.

\subsection{Requirements for planetary mass -- stellar metallicity investigations}\label{sec:required_samplemassmetallicity}

Extending the planetary mass -- stellar metallicity plane into the stellar mass dimension for GEMS requires not just additional planet discoveries, but also precise stellar metallicities for M-dwarfs. Furthermore, it will be important to determine the metallicities across the sample in a homogeneous manner to avoid methodology dependent metallicity offsets \citep{passegger_metallicities_2022}.


As can be seen in \autoref{fig:gems_masses}, the current sample of GEMS typically range in mass from around 0.3 -- 3 $M_J$. These will be further augmented through our follow-up of transiting GEMS, and also ongoing uninformed RV surveys with new instruments such as HPF \citep{mahadevan_habitable-zone_2014}, NEID \citep{halverson_comprehensive_2016, gupta_target_2021}, ESPRESSO \citep{pepe_espresso_2021}, CARMENES \citep{sabotta_carmenes_2021}, SPIROU \citep{donati_spirou_2020}, NIRPS \citep{wildi_nirps_2017}, IRD \citep{kotani_infrared_2018}, etc. Most importantly, Gaia astrometric detections (primarily expected in DR4 in 2025--2026\footnote{\url{https://www.cosmos.esa.int/web/gaia/release}}) are expected to add hundreds of GEMS and brown-dwarfs at intermediate orbital separations ($\sim$ AU) with mass measurements \citep{casertano_double-blind_2008, sozzetti_astrometric_2014, perryman_astrometric_2014}. An early example of this is GJ 463 b, for which a true mass was determined using a combination of RVs and Gaia astrometry \citep{endl_jupiter_2022, sozzetti_dynamical_2023}. This would offer a homogeneous and well-characterized sample at intermediate-long periods, that is ripe for population studies.

\section{Summary}\label{sec:summary}

In this manuscript we discuss the small but growing sample of transiting giant exoplanets around M-dwarf stars (GEMS), and the challenges posed to their formation by existing theories of planet formation by core-accretion and gravitational instability. We motivate the \textit{Searching for GEMS} survey to find, and characterize short-period transiting GEMS from TESS and ground-based facilities. We utilize multi-dimensional nonparametric statistics in the publicly available package \texttt{MRExo} to predict the required sample-size to robustly confirm the tentative trends (between the mass of Jovian-sized planets and host stellar mass) seen in the data to be about 40 confirmed transiting planets with mass measurements.  Lastly, we discuss the limitations of existing occurrence rates estimates for these GEMS, and highlight the stellar sample of $\sim$ 1 million M-dwarfs over a volume of 200 pc that we will use to characterize their occurrence.

\newpage
\section{Acknowledgement}

We thank the anonymous referee for the valuable feedback which has improved the quality of this manuscript.

S.K. acknowledges and appreciates discussions with Kevin Schlaufman regarding target selection and the Gaia queries, and also Theodora for proof-reading, and Annie Clark for providing a suitable background. GS acknowledges support provided by NASA through the NASA Hubble Fellowship grant HST-HF2-51519.001-A awarded by the Space Telescope Science Institute, which is operated by the Association of Universities for Research in Astronomy, Inc., for NASA, under contract NAS5-26555. CIC acknowledges support from an appointment to the NASA Postdoctoral Program at the Goddard Space Flight Center, administered by the ORAU through a contract with NASA.

These results are based on observations obtained with the Habitable-zone Planet Finder Spectrograph on the HET. We acknowledge support from NSF grants AST-1006676, AST-1126413, AST-1310885, AST-1310875,  ATI 2009889, ATI-2009982, AST-2108512, AST-2108801 and the NASA Astrobiology Institute (NNA09DA76A) in the pursuit of precision radial velocities in the NIR. The HPF team also acknowledges support from the Heising-Simons Foundation via grant 2017-0494. The Low Resolution Spectrograph 2 (LRS2) was developed and funded by the University of Texas at Austin McDonald Observatory and Department of Astronomy and by Pennsylvania State University. We thank the Leibniz-Institut für Astrophysik Potsdam (AIP) and the Institut für Astrophysik Göttingen (IAG) for their contributions to the construction of the integral field units.  The Hobby-Eberly Telescope is a joint project of the University of Texas at Austin, the Pennsylvania State University, Ludwig-Maximilians-Universität München, and Georg-August Universität Gottingen. The HET is named in honor of its principal benefactors, William P. Hobby and Robert E. Eberly. The HET collaboration acknowledges the support and resources from the Texas Advanced Computing Center. We thank the Resident astronomers and Telescope Operators at the HET for the skillful execution of our observations with HPF. We would like to acknowledge that the HET is built on Indigenous land. Moreover, we would like to acknowledge and pay our respects to the Carrizo \& Comecrudo, Coahuiltecan, Caddo, Tonkawa, Comanche, Lipan Apache, Alabama-Coushatta, Kickapoo, Tigua Pueblo, and all the American Indian and Indigenous Peoples and communities who have been or have become a part of these lands and territories in Texas, here on Turtle Island.

We acknowledge support from NSF grants AST-1910954, AST-1907622, AST-1909506, AST-1909682 for the ultra-precise photometry effort.

WIYN is a joint facility of the University of Wisconsin-Madison, Indiana University, NSF's NOIRLab, the Pennsylvania State University, Purdue University, University of California-Irvine, and the University of Missouri. 
The authors are honored to be permitted to conduct astronomical research on Iolkam Du'ag (Kitt Peak), a mountain with particular significance to the Tohono O'odham. Data presented herein were obtained at the WIYN Observatory from telescope time allocated to NN-EXPLORE through the scientific partnership of NASA, the NSF, and NOIRLab.

Data presented herein were obtained at the WIYN Observatory from telescope time allocated to NN-EXPLORE through the scientific partnership of the National Aeronautics and Space Administration, the National Science Foundation, and NOIRLab. This work was supported by a NASA WIYN PI Data Award, administered by the NASA Exoplanet Science Institute. These results are based on observations obtained with NEID on the WIYN 3.5 m telescope at KPNO, NSF's NOIRLab under proposal 2022B-785506 (PI: S. Kanodia), managed by the Association of Universities for Research in Astronomy (AURA) under a cooperative agreement with the NSF. This work was performed for the Jet Propulsion Laboratory, California Institute of Technology, sponsored by the United States Government under the Prime Contract 80NM0018D0004 between Caltech and NASA.

\software{
\texttt{astropy} \citep{robitaille_astropy_2013, astropy_collaboration_astropy_2018},
\texttt{ipython} \citep{perez_ipython_2007},
\texttt{matplotlib} \citep{hunter_matplotlib_2007},
\texttt{MRExo} \citep{kanodia_mass-radius_2019, kanodia_beyond_2023},
\texttt{numpy} \citep{oliphant_numpy_2006},
\texttt{pandas} \citep{mckinney_data_2010},
\texttt{scipy} \citep{oliphant_python_2007, virtanen_scipy_2020},
\texttt{tess-point} \citep{tesspoint},
}

\bibliography{references, manualreferences}




\end{document}